\documentclass[preprint,preprintnumbers,amsmath,amssymb,superscriptaddress]{revtex4}

\usepackage{graphicx,color}
\usepackage{amsmath,amssymb}
\usepackage{url}
\usepackage{epstopdf}
\usepackage{bbold}

\newcommand{\hs}{\hspace*{0.5cm}}

\newcommand{\be}{\begin{equation}}
\newcommand{\ee}{\end{equation}}
\newcommand{\bea}{\begin{eqnarray}}
\newcommand{\eea}{\end{eqnarray}}

\newcommand{\crn}{\nonumber \\}

\newcommand{\al}{\alpha}
\newcommand{\la}{\lambda}

\newcommand{\ga}{\gamma}

\newcommand{\fr}{\frac}
\newcommand{\bc}{\begin{center}}
\newcommand{\ec}{\end{center}}

\newcommand {\ba}{\begin{array}}
\newcommand {\ea}{\end{array}}
\newcommand{\ben}{\begin{enumerate}}
\newcommand{\een}{\end{enumerate}}

\usepackage{epsfig,graphicx}
\usepackage{bm}
\usepackage{dcolumn}
\allowdisplaybreaks
\begin{document}

\preprint{}

\title{Large signal of $h \rightarrow \mu \tau$ within the constraints of $e_i \rightarrow e_j\gamma$ decays in the 3-3-1 model with neutral leptons}
\author{H.~T.~Hung \footnote{Corresponding author}}\email{hathanhhung@hpu2.edu.vn}
\affiliation{Department of Physics, Hanoi Pedagogical University 2, Phuc Yen,  Vinh Phuc 15000, Vietnam}
\author{D. T. Binh}\email{dinhthanhbinh3@duytan.edu.vn}
\affiliation{Institute of Theoretical and Applied Research, Duy Tan University, Hanoi 10000, Vietnam
Faculty of Natural Science, Duy Tan University, Da Nang 50000, Vietnam.}
\author{H.~V.~Quyet}\email{hoangvanquyet@hpu2.edu.vn}
\affiliation{Department of Physics, Hanoi Pedagogical University 2, Phuc Yen,  Vinh Phuc 15000, Vietnam}

\begin{abstract}
In the framework of the 3-3-1 model with neutral leptons, we have investigated the lepton-flavor-violating sources based on the Higgs mass spectrum which has two neutral Higgses identitied with corresponding ones in the Two-Higgs-Doublet model (THDM). On the $13~\mathrm{TeV}$ scale of the LHC, we point out the parameter space regions where the experimental limits of $e_i \rightarrow e_j\gamma$ decays are satisfied. These regions depend heavily on the mixing of exotic leptons but are predicted to have large $h^0_1\rightarrow \mu \tau$ signals. We also show that $\mathrm{Br}(h^0_1\rightarrow \mu \tau)$ can reach a value of $10^{-4}$.
\end{abstract}

\pacs{{\bf Last updated: \today}
 }
\maketitle
\section{Introduction}

\allowdisplaybreaks
 The experimental data have confirmed the existence of oscillations of flavor neutrinos through the precise values of their mixing angles and their masses squared deviation \cite{Zyla:2020zbs}. This is an important base to assume about lepton-flavor-violating (LFV) decays. There are two types of LFV decays that have received a lot of attention, lepton-flavor-violating decays of charged leptons (cLFV) and lepton-flavor-violating decays of the standard model- like Higgs boson (LFVHDs). These LFV decays are concerned in both theory and experiment.  On the experimental side, cLFV are constrained by upper bounds as given in Ref.\cite{Patrignani:2016xqp},
\bea
Br(\mu \rightarrow e\gamma)<4.2\times 10^{-13},\crn
Br(\tau \rightarrow e\gamma)<3.3\times 10^{-8},\crn
Br(\tau \rightarrow \mu\gamma)<4.4\times 10^{-8}.\label{lalb-limmit}
\eea
These are currently the most stringent experimental limits for cLFV decays. It should be recalled that in addition to the cLFV limits given at Eq.(\ref{lalb-limmit}), we are interested in two other decay processes such as $\mu \rightarrow 3e$ and $\mu \rightarrow e$ conversion in nuclei. However, we also have experimental limits of these decay processes, $Br(\mu \rightarrow 3e)<10^{-12}$ Ref.\cite{SINDRUM:1987nra} and $CR(\mu^-Ti \rightarrow e^- Ti)<6.1\times 10^{-13}$ Ref.\cite{Lindner:2016bgg}, respectively. These limits are considered to be looser than that come from $Br(\mu \rightarrow e\gamma)$. Therefore, the limit of $Br(\mu \rightarrow e\gamma)$ can be used to find parameter space regions for the relevant process. Charged lepton flavor violation is considered to be a specific expression of the new physics that we are looking for. The hypothesis of its participation in the decays of heavy particles such as Z boson, Higgs boson or top quark is discussed in detail in Ref.\cite{Altmannshofer:2022fvz}.

For LFVHDs, the experimental limits are given as Refs.~\cite{CMS:2015qee,ATLAS:2015cji,CMS:2018ipm,ATLAS:2019erb,ATLAS:2019pmk},
\bea
Br(h \rightarrow \mu\tau)\leq \mathcal{O}(10^{-3}),\crn
Br(h \rightarrow \tau e)\leq \mathcal{O}(10^{-3}),\crn
Br(h \rightarrow \mu e)< 3.5 \times 10^{-4}. \label{hmt-limmit}
\eea
Then there is an adjustment $Br(h \rightarrow \mu e)< 6.1 \times 10^{-5}$  according to the update of Ref.~\cite{ATLAS:2019old}.

On the theoretical side, although LFV processes in general can receive tree and/or loop contributions, the LFV processes we study in this paper come only from loop diagrams. We therefore pay attention  both fermion and boson contributions. The fermions mentioned here include ordinary charged leptons, exotic leptons and neutrinos, but ordinary charged leptons are assumed to be unmixed so its contribution can be determined relatively simply. The complex part belongs to neutrinos and exotic leptons with different mixing mechanisms. With active neutrinos, we can solve their masses and oscillations using seesaw mechanisms Refs.~\cite{Gomez:2017dhl,CarcamoHernandez:2019pmy,Catano:2012kw,Hernandez:2014lpa,Dias:2012xp,Nguyen:2018rlb,Hue:2021zyw} or otherwise Refs.~\cite{Hue:2017lak,Thuc:2016qva,Marcano:2019rmk, Das:2020pai}, with exotic leptons, we have different assumptions for large LFV effects as given in  Refs.~\cite{Hue:2017lak,Thuc:2016qva,Hong:2020qxc}. On the contribution of bosons, we pay attention to both the charged gauge bosons and the charged Higgses. The main contribution of the gauge bosons comes from the new charged bosons, which are outside the standard model, because the contribution of the W-boson is strongly suppressed by GIM mechanism. The contributions of charged Higgses are varied and depend heavily on the energy scales of the accelerators.

It should be emphasized that LFV sources mainly come from the models beyond the standard model (BSM), and we are interested in the  parameter space domains predicted from BSM for the large  signal of LFVHDs is limited directly from both the experimental data and theory of cLFV \cite{Herrero-Garcia:2016uab,Blankenburg:2012ex}. Some published results show that $Br(h_1^0\rightarrow \mu\tau)$ can reach values of $\mathcal{O}(10 ^ {-4})$ in supersymmetric and non-supersymmetric models \cite{Zhang:2015csm,Herrero-Garcia:2017xdu,Qin:2017aju}. In addition to the correction from the loop, other ways are also suggested in the literature for large $h \rightarrow \mu \tau$ signal. For example, by using the type-I seesaw mechanism and an effective dimension-six operator, one is possible to accommodate the CMS $h \rightarrow \mu \tau$ signal with a branching ratio of order $10^{-2}$ \cite{He:2015rqa}. In fact, the main contributions to $Br(h_1^0\rightarrow \mu\tau)$ come from new heavy particles in BSM. If those contributions are minor or destructive, the $Br(h_1^0\rightarrow \mu\tau)$ in a model is only about $\mathcal{O}(10 ^ {-9})$ \cite{Gomez:2017dhl}. 

Recently, the 3-3-1 models with multiple sources of lepton flavor violating couplings are used to investigate LFV decays \cite{PhysRevD.22.738,Chang:2006aa, Okada:2016whh,  Dong:2008sw,  Dias:2006ns, Diaz:2004fs, Diaz:2003dk, Fonseca:2016xsy, Buras:2012dp, Buras:2014yna}. 
However, these models can only give very small LFV signals or cLFV and LFVHDs can achieve relatively large signals but in different regions of the parameter space \cite{Hue:2015fbb, Thuc:2016qva,Boucenna:2015zwa,Hernandez:2013hea}. In particular, detailed calculations of cLFV are given in Ref.\cite{Hue:2017lak} without mentioning LFVHDs, whereas a 3-3-1 model mentioned in Ref.\cite{Hue:2015fbb} only examines LFVHDs. Several other versions of the 3-3-1 models have used the inverse seesaw mechanism to study the LFV decays. In this way, it is necessary to introduce new particles that are singlets of gauge group, leading to an increase in the number of particles as well as free parameters in those models \cite{Nguyen:2018rlb}. The 3-3-1 model with neutral leptons can reduce the number of free parameters because without the heavy particles as singlets of gauge group, the LFV source comes from the usual mixing of neutrinos and neutral leptons. This is a good model for studying both cLFV and LFVHDs. Besides the LFV decays, the 3-3-1 models can also give large signals of other SM-like Higgs boson decays such as, $h^0_1\rightarrow \gamma \gamma$ and $h^0_1\rightarrow Z \gamma$ \cite{Hung:2019jue,Phan:2021pcc} .

In this work, we will consider a 3-3-1 model to find regions of the parameter space that satisfy the experimental limits of cLFV. In these regions, we predict the existence of a large signal of $h^0_1\rightarrow \mu \tau$ decay. Combined with the signal of $h^0_1\rightarrow Z \gamma$ as given in Ref.~\cite{Hung:2019jue}, we expect to have the parameter space regions for large signals of both $h^0_1\rightarrow \mu \tau$ and $h^0_1\rightarrow Z \gamma$ decays.
 
The paper is organized as follows. In the next section, we review the model and give masses spectrum of gauge and Higgs bosons. We then show the masses spectrum of the all leptons in Section \ref{Couplings}. We calculate the Feynman rules and analytic formulas for cLFV and LFVHDs  in Section \ref{Analytic}. Numerical results are discussed in Section.\ref{numerical_results}. Conclusions are in Section \ref{conclusion}. Finally, we provide Appendix \ref{appen_PV}, \ref{appen_loops1}, \ref{appen_loops2} to calculate and exclude divergence in the amplitude of LFVHDs.

\section{ The review of 3-3-1 model with neutral leptons}
\label{331LHN}
The 3-3-1 model model with neutral leptons is a specific class of general 3-3-1 models (331$\beta$), which obey the gauge symmetry group $SU(3)_C\otimes SU(3)_L\otimes U(1)_X$ and the parameter $\beta=-\frac{1}{\sqrt{3}}$. The parameter $\beta$ is a basis for defining the form of electric charge operator in this model: $Q=T_3+\beta T_8+X$, where $T_{3,8}$ are diagonal $SU(3)_L$ generators. The model under consideration is developed based on the following highlights: {\it{i})} the active neutrinos have no right-handed components, so they have only Majorana masses which are generated from the effective dimension-five operators and there is no mixing among active neutrinos and exotic leptons Ref.~\cite{Mizukoshi:2010ky}, {\it{ii})} exotic leptons are always assumed to have large mixing for the appearance of the LFV effect Ref.~\cite{Hue:2015fbb}, {\it{iii})} there are two neutral Higgs that are identified with the corresponding ones of the THDM, with the expectation of having both large signals of $h_1^0\rightarrow \mu\tau$ and $h_1^0\rightarrow Z\gamma$ decays. Thus, we will call this model 331NL for short form.

 The leptons in the 331NL model are accommodated in triplets and singlets representations as follows: 

\begin{eqnarray}
\Psi_{aL}^\prime = \left (
\begin{array}{c}
\nu^\prime_{a} \\
e^\prime_{a} \\
N^\prime_{a}
\end{array}
\right )_L\sim(1\,,\,3\,,\,-1/3)\,,\,\,\,e^\prime_{aR}\,\sim(1,1,-1)\,,\,\,\,N^\prime_{aR}\,\sim(1,1,0),
\label{L}
\end{eqnarray}
where $a=1,2,3$ represents the family index for the usual three generation of leptons, the numbers in the parentheses are  the respective representations of the $SU(3)_C$, $SU(3)_L$  and $U(1)_X$ gauge groups. We use the primes to denote for the fermions in the flavor basis. The right-handed components of the charged leptons and the exotic neutral leptons are $e^\prime_{aR}$ and $N^\prime_{aR}$, respectively. $N^\prime_{aL,R}$ are also the new degrees of freedom in the model.

In the quark sector, the third generation comes in the triplet representation and the other two are in an anti-triplet representation of $SU(3)_L$, as a requirement for anomaly cancellation. They are given by,
\begin{eqnarray}
&&Q_{iL}^\prime = \left (
\begin{array}{c}
d^{\prime}_{i} \\
-u^{\prime}_{i} \\
D^{\prime}_{i}
\end{array}
\right )_L\sim(3\,,\,\bar{3}\,,\,0)\,, \nonumber \\
&&
u^{\prime}_{iR}\,\sim(3,1,2/3),\,\,\,
\,\,d^{\prime}_{iR}\,\sim(3,1,-1/3)\,,\,\,\,\, D^{\prime}_{iR}\,\sim(3,1,-1/3),\nonumber \\
&&Q_{3L}^\prime = \left (
\begin{array}{c}
u^{\prime}_{3} \\
d^{\prime}_{3} \\
U^{\prime}_{3}
\end{array}
\right )_L\sim(3\,,\,3\,,\,1/3)\,, \nonumber \\
&&
u^{\prime}_{3R}\,\sim(3,1,2/3),
\,\,d^{\prime}_{3R}\,\sim(3,1,-1/3)\,,\,U^{\prime}_{3R}\,\sim(3,1,2/3)
\label{quarks} 
\end{eqnarray}
where the index $i=1,2$ was chosen to represent the first two generations. $U^{\prime}_{3L,R}$ and $D^{\prime}_{iL,R}$ are new heavy quarks with the usual fractional electric charges.

The scalar part is introduced three triplets, which are guaranteed to generate the masses of the SM fermions, 
\begin{eqnarray}
\eta = \left (
\begin{array}{c}
\eta^0 \\
\eta^- \\
\eta^{\prime 0}
\end{array}
\right ),\,\rho = \left (
\begin{array}{c}
\rho^+ \\
\rho^0 \\
\rho^{\prime +}
\end{array}
\right ),\,
\chi = \left (
\begin{array}{c}
\chi^{\prime 0} \\
\chi^{-} \\
\chi^0
\end{array}
\right )\,, 
\label{scalarcont} 
\end{eqnarray}
with $\eta$ and $\chi$ both transforming as $(1\,,\,3\,,\,-1/3)$
and $\rho$ transforming as $(1\,,\,3\,,\,2/3)$.

The 331NL model exits two global symmetries, namely $L$ and $\mathcal{L}$ are the normal and new lepton numbers, respectively Refs.~\cite{Chang:2006aa,Tully:2000kk}. They are related to each other by $L=\frac{4}{\sqrt{3}}T_8+\mathcal{L}$ with $T_8=\frac{1}{2\sqrt{3}}\mathrm{diag}(1,1,-2)$. So, $L$ and $\mathcal{L}$ of multiplet in the model are given as,

\begin{table}[h]
	\begin{tabular}{|c|ccc|cccccc|ccc|}
		\hline
		Multiplet & $\Psi_{aL}^\prime$ & $e_{aR}^{\prime}$ & $N_{aR}^{\prime}$& $Q_{i L}^{\prime}$ & $Q_{3L}^{\prime}$& $u_{aR}^{\prime}$& $d_{aR}^{\prime}$&  $D_{iR}^{\prime}$ & $U_{3R}^{\prime}$ & $\eta$ & $\rho$ & $\chi$ \\
		\hline
		$\mathcal{L}$ & $\frac{1}{3}$ & $1$ & $1$&  $\frac{2}{3}$ & $-\frac{1}{3}$& $0$ & $0$ & $2$ & $-2$ & $-\frac{2}{3}$ & $-\frac{2}{3}$ & $\frac{4}{3}$ \\
		\hline
	\end{tabular}
	\caption{The lepton numbers $\mathcal{L}$ of all multiplet in the 331NL model.} \label{L_number}
\end{table} 

The number $L$ assigned to each field is

\begin{table}[h]
		\begin{tabular}{|c|ccccc|cccc|ccccccccc|}
			\hline
			Fields & $\nu_{aL}^\prime$ & $e_{aL}^{\prime}$ & $N_{aL}^{\prime}$ & $e_{aR}^{\prime}$ & $N_{aR}^{\prime}$& $u_{a L,R}^{\prime}$ & $d_{aL,R}^{\prime}$ & $D_{i L,R}^{\prime}$ & $U_{3L,R}^{\prime}$& $\eta^{-}$& $\eta^{0}$&  $\eta^{\prime 0}$ & $\rho^{+}$ & $\rho^0$ & $\rho^{\prime +}$ & $\chi^{\prime 0}$ & $\chi^{-}$ & $\chi^{0}$ \\
			\hline
			$L$ & $1$ & $1$ & $-1$&  $1$ & $1$ & $0$ & $0$ & $2$ & $-2$ & $0$ & $0$ & $-2$ & $0$ & $0$ & $-2$ & $2$ & $2$ & $0$  \\
			\hline
		\end{tabular}\caption{The lepton numbers $L$ of the fields in the 331NL model.} \label{zero_vev}
\end{table} 

 As a result, the normal lepton number $L$ of $\eta^0$, $\rho^0$ and $\chi^0$ are zeros. In contrast, $\eta'^0$ and $\chi'^0$ are bilepton with $L=\mp 2$. This is the difference in lepton numbers of the components in the $\eta$ and $\chi$ triplets. To break $SU(3)_L$, we need the vacuum expectation values (VEV) $\left\langle \chi^0\right\rangle $ to be non-zero and in scale of exotic particle masses. Thus, one can convention $\left\langle \eta^{\prime 0}\right\rangle $ to be zero. From Eq.(\ref{quarks}), the generations have different gauge charge so we need  $\eta, \rho$ triplets to break $SU(2)_L$. Mean, we require non-zero $\left\langle \eta^0 \right\rangle $ and $\left\langle \rho^0 \right\rangle $ to ensure that condition, then $\left\langle \chi^{\prime 0}\right\rangle $ can be chosen to be zero to reduce the free parameter in the model. 

Thus, all VEVs in this model are introduced as follow, 
\bea \eta^{\prime 0}&=& \frac{S'_2+i A'_2}{\sqrt{2}},\hs \chi^{\prime 0}= \frac{S'_3+i A'_3}{\sqrt{2}} \crn
\rho^0 &=&\frac{1}{\sqrt{2}}\left(v_1+S_1+iA_1\right),\; \eta^0=\frac{1}{\sqrt{2}}\left(v_2+S_2+iA_2\right),\; \chi^0=\frac{1}{\sqrt{2}}\left(v_3+S_3+iA_3\right). \label{vevs}\eea
The electroweak symmetry breaking (EWSB) mechanism follows
\begin{equation*}
{SU(3)_{L}\otimes U(1)_{X}\xrightarrow{\langle \chi \rangle}}{%
	SU(2)_{L}\otimes U(1)_{Y}}{\xrightarrow{\langle \eta \rangle,\langle \rho
		\rangle}}{U(1)_{Q}},
\end{equation*}
where VEVs satisfy the hierarchy ${v_3 \gg v_1,v_2}$  as done
in Refs. \cite{Dong:2008sw,Dong:2010gk}. 

The most general scalar potential was constructed based on Refs.\cite{Chang:2006aa,Hue:2021xap} has the form,
\begin{eqnarray} V(\eta,\rho,\chi)&=&\mu_1^2\eta^2
+\mu_2^2\rho^2+\mu_3^2 \chi^2 +\lambda_1\eta^4
+\lambda_2\rho^4+\lambda_3\chi^4  \nonumber \\
&&+\lambda_{12}
(\eta^{\dagger}\eta)(\rho^{\dagger}\rho)+\lambda_{13}(\eta^{\dagger}\eta)(\chi^{\dagger}\chi)
+\lambda_{23}(\rho^{\dagger}\rho)(\chi^{\dagger}\chi) \nonumber \\
&&+\tilde{\lambda}_{12}
(\eta^{\dagger}\rho)(\rho^{\dagger}\eta)+\tilde{\lambda}_{13}(\eta^{\dagger}\chi)(\chi^{\dagger}\eta)
+\tilde{\lambda}_{23}(\rho^{\dagger}\chi)(\chi^{\dagger}\rho) \nonumber \\
&&+\sqrt{2}fv_3\left( \epsilon^{ijk}\eta_i \rho_j \chi_k +\mbox{H.c}\right).
\label{potential}
\end{eqnarray}
where $f$ is a dimensionless coefficient that is included for convenience in later calculations. Compared to the general form in Ref.\cite{Chang:2006aa}, small terms in the Higgs potential in Eq.~(\ref{potential}) that violating the lepton number have been ignored. But it still gives this model a diverse Higgs mass spectrum. The masses and physical states of Higgs bosons and gauge bosons are given in App.\ref{appen_HGB}.

\section{Couplings for LFV decays}
\label{Couplings}

 We use the Yukawa terms shown in Ref.~\cite{Mizukoshi:2010ky} for generating masses of charged leptons, active neutrinos and exotic neutral leptons, namely
\be -\mathcal{L}^{Y}_{\mathrm{lepton}} = h^{e}_{ab}\overline{\Psi^\prime_{a}}\rho e'_{bR}+ h^{N}_{ab}\overline{\Psi^\prime_{a}}\chi N'_{bR}+ \frac{h^{\nu}_{ab}}{\Lambda} \left(\overline{(\Psi^\prime_{a})^c}\eta^*\right)\left(\eta^{\dagger}\Psi^\prime_{b}\right) + \mathrm{h.c.}, \label{Ylepton1}\ee
where the notation $(\Psi^\prime)^c_a=( (\nu'_{aL})^c,\;(e'_{aL})^c,\;(N'_{aL})^c\;)^T\equiv( \nu'^c_{aR},\;e'^c_{aR},\;N'^c_{aR}\;)^T$  implies that $\psi^c_R\equiv P_R \psi^c= (\psi_L)^c$ with $\psi$ and $\psi^c \equiv C\overline{\psi}^T$ being the Dirac spinor and its charge conjugation, respectively. Remind that $P_{R,L}\equiv \frac{1\pm\gamma_5}{2}$ are the right- and left-chiral projection operators, we have $\psi_L= P_L \psi, \; \psi_R=P_R\psi$. The $\Lambda$ is some high energy scale.
The corresponding mass terms are
\be -\mathcal{L}^{mass}_{\mathrm{lepton}} = \left[ \frac{h^{e}_{ab}v_1}{\sqrt{2}}\overline{e'_{aL}} e'_{bR}+\frac{ h^{N}_{ab}v_3}{\sqrt{2}}\overline{N'_{aL}}  N'_{bR}+ \mathrm{h.c.} \right]+ \frac{h^{\nu}_{ab}v^2_2}{2 \Lambda}\left[ (\overline{\nu'^c_{aR}} \nu'_{bL})+ \mathrm{h.c.}\right]. \label{mterm}\ee
Since there are no right-handed components, active neutrinos have only Majorana masses. Their mass matrix is  $(M_{\nu})_{ab} \equiv  \frac{h^{\nu}_{ab}v^2_2}{ \Lambda}$. and proved to be symmetric based on Ref.~\cite{Mohapatra:1991ng}, therefore the mass eigenstates can be found by a single rotation expressed by a  mixing matrix $U$ that satisfies $ U^{\dagger} M_{\nu}U=\mathrm{diagonal}(m_{\nu_1},\;m_{\nu_2}, \;m_{\nu_3})$, where $m_{\nu_i}$ (i=1,2,3) are mass eigenvalues of the active neutrinos.

We now define transformations between the flavor basis  $\{e'_{aL,R},~\nu'_{aL},~N'_{aL,R}\}$ and the mass basis $\{e_{aL,R},~\nu_{aL},~N_{aL,R}\}$:
\be
e'^-_{aL}= e^-_{aL},  ~~e'^-_{aR}=e^-_{aR}, \;
\nu'_{aL}=U_{ab}\nu_{bL},\; N'_{aL}=V^L_{ab}N_{bL},\quad N'_{aR}=V^R_{ab}N_{bR},
\label{lepmixing}\ee
where $V^L_{ab},~U^L_{ab}$ and  $V^R_{ab}$ are transformations between flavor and mass bases of  leptons. Here, primed fields and unprimed fields denote the flavor basis and the mass eigenstates, respectively. Denote that $\nu'^c_{aR}=(\nu'_{aL})^c=U_{ab}\nu^c_{aR}$. The four-spinors representing the active neutrinos are $\nu^c_{a}=\nu_{a}\equiv (\nu_{aL},\; \nu^c_{aR})^T$, resulting the following equalities: $\nu_{aL}=P_L\nu^c_a=P_L\nu_a$ and $\nu^c_{aR}=P_R\nu^c_a=P_R\nu_a$. Experiments have not yet found the oscillation of charged leptons. This is confirmed again in Refs.\cite{BaBar:2009hkt,Hayasaka:2010np,MEG:2011naj}. As the results, the upper bounds of recent experiments for the LFV processes in the normal charged  leptons are very suppressed, therefore imply that the two flavor and mass bases  of charged leptons should be the same.

The relations between the mass matrices of leptons in  two flavor and mass bases are
\bea m_{e_a}&=&\frac{v_1}{\sqrt{2}}h^e_{a},\hs h^e_{ab}=h^e_a\delta_{ab},\hs a,b=1,2,3,
\crn \frac{v_2^2}{\Lambda} U^{\dagger}H^{\nu} U&=& \mathrm{Diagonal}(m_{\nu_1},~m_{\nu_2},~m_{\nu_3}),\crn
\frac{v_3}{\sqrt{2}} V^{L\dagger}H^N V^R&=& \mathrm{Diagonal}(m_{N_1},~m_{N_2},~m_{N_3}),\label{cema1} \eea
where $H^{\nu}$ and $H^N$ are Yukawa matrices defined as $(H^{\nu})_{ab}=h^{\nu}_{ab}$ and $(H^{N})_{ab}=h^{N}_{ab}$.

The  Yukawa interactions between leptons and Higgses can be written according to the lepton mass eigenstates,
{\small \bea  -\mathcal{L}^{Y}_{\mathrm{lepton}} &=&\frac{m_{e_b}}{v_1}\sqrt{2} \left[\rho^{0} \bar{e}_bP_Re_b+  U^{*}_{ba}\bar{\nu}_a P_Re_b\rho^{+} + V^{L*}_{ba}\overline{N}_a P_Re_b\rho'^{+}+\mathrm{h.c.}   \right]\crn
	&&+\frac{m_{N_a}}{v_3}\sqrt{2} \left[\chi^{0} \bar{N}_aP_RN_a+  V^{L}_{ba}\bar{e}_b P_RN_a\chi^{-}+\mathrm{h.c.}   \right]\crn
	&&+ \frac{m_{\nu_a}}{v_2}\left[S_2\overline{\nu_{a}}P_L\nu_{b}+\fr{1}{\sqrt{2}} \eta^{+}\left(U^{*}_{ba} \overline{\nu_{a}}P_Le_{b}+ U_{ba} \overline{e^c_{b}}P_L\nu_{a}\right)+\mathrm{h.c.} \right], \label{llh}\eea}
where we have used the Majorana property of the active neutrinos: $\nu^c_a=\nu_a$ with $a=1,2,3$.  In addition, using the equality $\overline{e^c_{b}}P_L\nu_{a}=  \overline{\nu_{a}}P_Le_{b}$ for this case the term relating with $\eta^{\pm}$ in the last line of (\ref{llh}) is reduced to $\sqrt{2}\eta^{+} \overline{\nu_{a}}P_Le_{b}$.

The covariant derivatives  of the leptons contain the lepton-lepton-gauge boson couplings, namely
\bea \mathcal{L}^D_{\mathrm{lepton}} &=& i\overline{L'_a}\gamma^{\mu}D_{\mu}L'_a\crn
&\rightarrow& \frac{g}{\sqrt{2}}\left[ U^*_{ba}\overline{\nu_a}\gamma^{\mu}P_L e_bW^+_{\mu} +U_{ab}\overline{e_b}\gamma^{\mu}P_L\nu_aW^-_{\mu}\right. \crn &+&\left. V^{L*}_{ba}\overline{N_a}\gamma^{\mu}P_L e_bV^+_{\mu} +V^L_{ab}\overline{e_b}\gamma^{\mu}P_LN_aV^-_{\mu}  \right] . \label{cdelepton}\eea
The couplings of the Higgses with the gauge bosons comes from the covariant derivative of the scalar fields.
\bea \mathcal{L}^D_{\mathrm{scalar}} = i\sum_{\Phi=\eta,\rho,\chi}\overline{\Phi}\gamma^{\mu}D_{\mu}\Phi. \label{coupHiggs-gauge}\eea
Based on Eq.(\ref{coupHiggs-gauge}), we obtain couplings of SM-like Higgs with charged gauge bosons and charged Higgses. In particular, regarding the interactions of charged Higgs with $W$-boson and $Z$-boson mentioned as Refs.\cite{ATLAS:2015edr, CMS:2015lsf}, we find out that in this model only $H_1^\pm W^\mp Z$ is non-zero and $H_2^\pm W^\mp Z$ is suppressed. This results in $m_{H_1^\pm}$ being limited to around $600~\mathrm{GeV}$ \cite{ATLAS:2015edr} or around $1.0~\mathrm{TeV}$ \cite{CMS:2015lsf}.

From the above expansions, we show the couplings relating to cLFV and LFVHDs of this model in Table. \ref{albga}.

\begin{table}[h]
	\scalebox{0.82}{
		\begin{tabular}{|c|c|c|c|}
			\hline
			Vertex & Coupling & Vertex &Coupling \\
			\hline
		$\bar{\nu}_ae_bH_1^{+}$&$-i\sqrt{2}U^{L*}_{ba}\left( \dfrac{m_{e_b}}{v_1}c_{12}P_R+\dfrac{m_{\nu_a}}{v_2}s_{12}P_L\right)$&$\bar{e}_b\nu_aH_1^{-}$&$-i\sqrt{2}U^{L}_{ab}\left( \dfrac{m_{e_b}}{v_1}c_{12}P_L+\dfrac{m_{\nu_a}}{v_2}s_{12}P_R\right)$\\
		\hline	
			$\bar{N}_a e_bH_2^{+}$ & $-i\sqrt{2}V^{L*}_{ba}\left(\fr{m_{e_b}}{v_1}c_{13} P_R+\fr{m_{N_a}}{v_3}s_{13} P_L\right)$ & $\bar{e}_a N_bH_2^{-}$ &$-i\sqrt{2}V^{L}_{ba}\left(\fr{m_{e_b}}{v_1}c_{13} P_L+\fr{m_{N_a}}{v_3}s_{13} P_R\right)$ \\
			\hline
			$\bar{e}_ae_ah_1^0$&$-\fr{im_{e_a}}{v_1}s_\al$&$\bar{\nu}_a\nu_ah_1^0$&$\fr{im_{\nu_a}c_\al}{v_2}$\\
			\hline
			$\bar{N}_ae_bV_\mu^{+}$&$\fr{ig}{\sqrt{2}}V^{L*}_{ba}\ga^\mu P_L$&$\bar{e}_bN_aV_\mu^{-}$&$\fr{ig}{\sqrt{2}}V^{L}_{ab}\ga^\mu P_L$\\
			\hline
			$\bar{\nu}_ae_bW_\mu^{+}$&$\fr{ig}{\sqrt{2}}U^{L*}_{ba}\ga^\mu P_L$&$\bar{e}_b\nu_aW_\mu^{-}$&$\fr{ig}{\sqrt{2}}U^{L}_{ab}\ga^\mu P_L$\\
			\hline
			$W^{\mu+}W_\mu^{-}h_1^0$&$\fr{ig}{2}m_W\left(  c_\al s_{12}- s_\al c_{12}\right)  $&$V^{\mu+}V_\mu^{-}h_1^0$&$-\dfrac{ig}{2}m_W s_\al c_{12}$\\
			\hline
			$h_1^0H_1^{+}W^{\mu-}$&$\dfrac{ig}{2}\left(c_\al c_{12}+s_\al s_{12} \right) (p_{h_1^0}-p_{H_1^{+}})_\mu$&$h_1^0H_1^{-}W^{\mu+}$&$\dfrac{ig}{2}\left(c_\al c_{12}+s_\al s_{12} \right)(p_{H_1^{-}}-p_{h_1^0})_\mu$ \\
			\hline
			$h_1^0H_2^{+}V^{\mu-}$&$\dfrac{ig}{2}s_\al c_{13}(p_{h_1^0}-p_{H_2^{+}})_\mu$&$h_1^0H_2^{-}V^{\mu+}$&$\dfrac{ig}{2}s_\al c_{13}(p_{H_2^{-}}-p_{h_1^0})_\mu$ \\
			\hline
			$h_1^0 H_1^{+}H_1^{-}$&$-i \lambda_{h^0H_1H_1} $&$h_1^0H_2^{+}H_2^{-}$&$ -i \lambda_{h^0H_2H_2}$\\
			\hline
	\end{tabular}}
	\caption{ The couplings relating to cLFV and LFVHDs in the 331NL model. All the couplings were only considered in the unitary gauge.} \label{albga}
\end{table}
The self-couplings of Higgs bosons are given as: 
\bea
\lambda_{h^0H_1H_1}&&=\left[ \left(c_{12}^3c_\al -s_{12}^3s_\al \right)  \left(\lambda_{12} +\tilde{\lambda}_{12}\right) -c_{12}^2 s_{12}^2 s_\al\left( 2\lambda_{2}+\tilde{\lambda}_{12}\right) +  s_{12}^2 c_{12}^2 c_\al\left( 2\lambda_{1}+\tilde{\lambda}_{12}\right) \right] \sqrt{ v_1^2+v_2^2},\crn
\lambda_{h^0H_2H_2}&&=\left[ c_{12}^2 s_{12}s_\al\left(\lambda_{23}+\tilde{\lambda}_{23}\right)-2c_{13}^3s_{12}s_\al \lambda_{2}+c_{13}^3c_{12}c_\al\lambda_{12}-s_{13}^3c_{12}c_\al\lambda_{13}\right] \sqrt{ v_1^2+v_2^2} \crn
&&+ c_{13}s_{13}\left(s_\al \tilde{\lambda}_{23}-2fc_\al \right)v_3.
\eea

In Tab.(\ref{albga}), we realize that the flavor-diagonal modes $h_1^0 \rightarrow e^+_a e^-_a$ occur naturally at the tree level. Because, the corresponding vertices are not suppressed, $h_1^0\bar{e}_ae_a =-\frac{im_{e_a}s_\alpha}{v_1}=\frac{im_{e_a}}{v}.\frac{c_{(\beta_{12}+\delta)}}{c_{\beta_{12}}}$. Recall that, we have $h^0\bar{e}_ae_a =\frac{im_{e_a}}{v}$ in the SM. Therefore, the $h_1^0\bar{e}_ae_a$ decays in this model are implemented in parameter domains different from the SM. This difference is determined through a coefficient $\frac{c_{(\beta_{12}+\delta)}}{c_{\beta_{12}}}$. It is very small and will suppress in the limit $\delta \rightarrow 0$.


\section{Analytic formulas for contributions to LFVHD and cLFV decays}
\label{Analytic}
This model has a striking resemblance to the standard model, in that the W-boson only couplings with active neutrinos. In contrast, exotic neutrinos coupling with both the newly charged gauge boson and the heavily charged Higgs. By putting on an align limit in Eq.(\ref{eq_alignH0}) and mixing of neutral Higgs in Eq.(\ref{mixingh0}), we obtain $h^0_1$ which fully inherits the same characteristics as SM-like Higgs in THDM that were shown in Ref.\cite{Okada:2016whh}. However, this also leads to the consequences that some couplings such as: $h^0_1\overline{N}_aN_a$, $h^0_1H_1^\pm H_2^\mp$, $h^0_1H_1^\pm V^\mp$, $h^0_1H_2^\pm W^\mp$ are canceled out. 
\subsection{\label{Analytic1} Analytic formulas for $e_i \rightarrow e_j\gamma$ decays} 
 In this section, we pay attention to one-loop order contributions of cLFV decays. Based on Tab.~\ref{albga}, all Feynman diagrams at one-loop order for $e_i \rightarrow e_j \gamma$ decays are given as shown below,
\begin{figure}[ht]
	\centering
	\begin{tabular}{cc}
		\includegraphics[width=14.0cm]{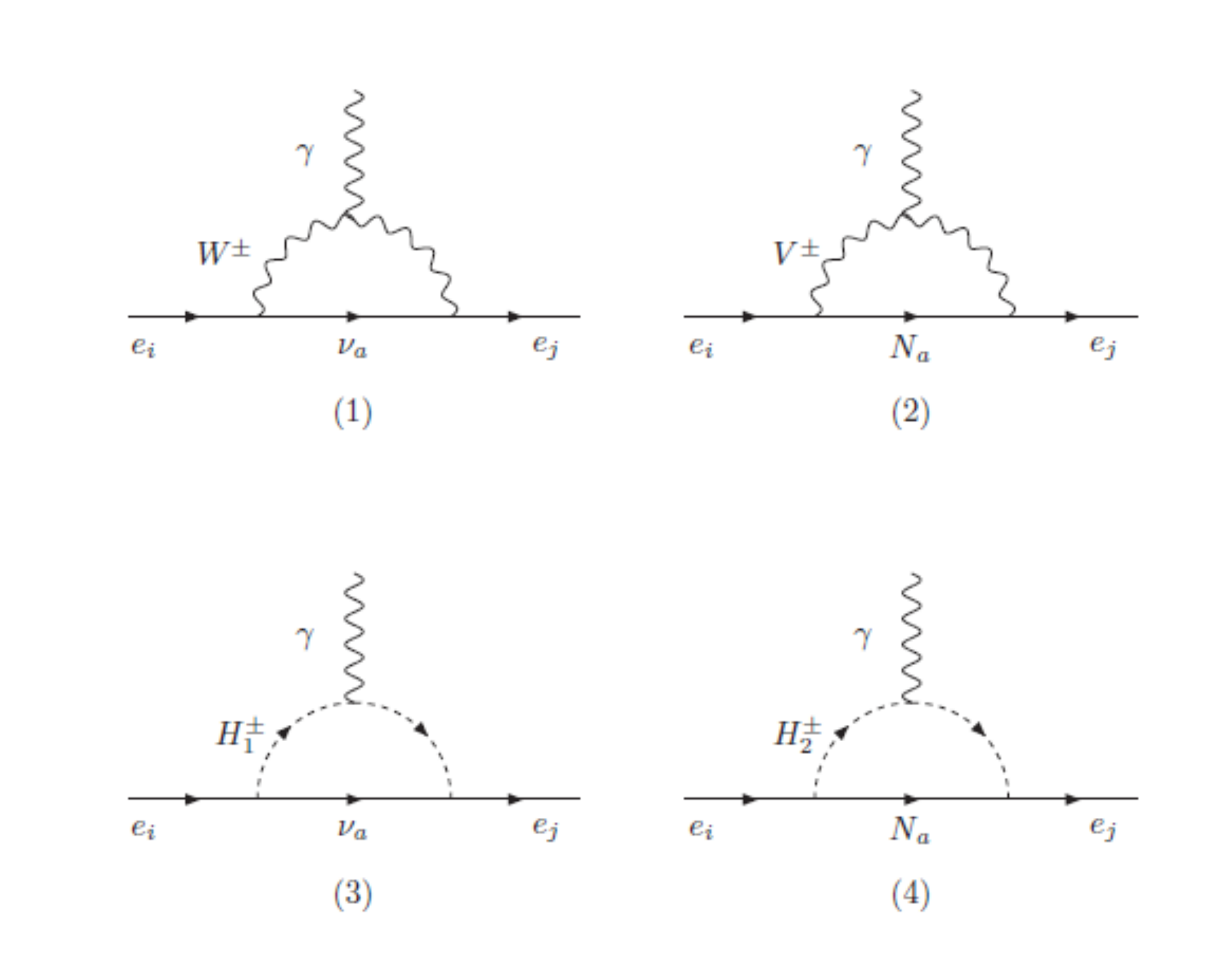} 
	\end{tabular}%
	\caption{ Feynman diagrams at one-loop order of $e_i \rightarrow e_j \gamma$ decays in the unitary gauge.}
	\label{fig_lalbga}
\end{figure}

The general form of cLFV decays is given as
\bea e_i(p_i) \rightarrow e_j(p_j) + \gamma (q),
\eea
where $p_i=p_j+q$.
The amplitude is known
\bea \mathcal{M}=\epsilon_{\lambda}\overline{u}_i(p_i)\Gamma^{\lambda} u_j(p_j),
\eea  
where $\epsilon_{\lambda}$ is the polarization vector of photon, $\Gamma^{\lambda}$ are $4\times4$ matrices depending on external momenta, coupling constants and the gamma matrices. By using formulas $\epsilon_{\mu}q^{\mu}=0$ and $q_{\lambda}\overline{u}_i(p_i)\Gamma^{\lambda} u_j(p_j)=0$, we can get the form of the amplitude as,
\bea
\mathcal{M}&&=\overline{u}_j(p_j) \left[2\left(p_i.\epsilon \right)\left(  \mathcal{C}_{(ij)L}P_L+\mathcal{C}_{(ij)R}P_R\right) \right.\crn
&&\left.-(m_i\mathcal{C}_{(ij)R}+m_j\mathcal{C}_{(ij)L})\epsilon\!\!/P_L-(m_i\mathcal{C}_{(ij)L}+m_j\mathcal{C}_{(ij)R})\epsilon \!\!/ P_R \right]u_i(p_i) 
\eea
where $P_L=\frac{1-\gamma_5}{2},\,P_R=\frac{1+\gamma_5}{2}$ and $\mathcal{C}_{(ij)L},\,\mathcal{C}_{(ij)R}$ are factors.

For the convenience of calculations, we denote: $\mathcal{C}_{(ij)L}=2m_j\mathcal{D}_{(ij)L}$ and $\mathcal{C}_{(ij)R}=2m_i\mathcal{D}_{(ij)R}$. 
Based on the discussions in Refs.~\cite{Hue:2017lak, Crivellin:2018qmi}, we can get the total branching ratios   of the cLFV  processes as
\begin{equation}\label{eq_Gaebaga}
\mathrm{Br}^{Total}(e_i\rightarrow e_j\gamma)\simeq \frac{48\pi^2}{ G_F^2} \left( \left|\mathcal{D}_{(ij)R}\right|^2 +\left|\mathcal{D}_{(ji)L}\right|^2\right) \mathrm{Br}(e_i\rightarrow e_j\overline{\nu_j}\nu_i),
\end{equation}
where $G_F=g^2/(4\sqrt{2}m_W^2)$, and for different charge lepton decays, we use experimental data $\mathrm{Br}(\mu\rightarrow e\overline{\nu_e}\nu_\mu)=100\%, \mathrm{Br}(\tau\rightarrow e\overline{\nu_e}\nu_\tau)=17.82\%, \mathrm{Br}(\tau\rightarrow \mu\overline{\nu_\mu}\nu_\tau)=17.39\% $ as given in Ref.\cite{Patrignani:2016xqp,Tanabashi:2018oca,Zyla:2020zbs}. 
This result is consistent with the  formulas given used in Refs.~\cite{Hue:2017lak, Nguyen:2018rlb,Hung:2021fzb,Hue:2021zyw,Hue:2021xap,Hong:2020qxc} for 3-3-1 models.

Analytical results of the diagrams in Fig.\ref{fig_lalbga} are given in Appendix \ref{appen_loops1}.
The total one-loop contribution to the cLFV decays $e_i\rightarrow e_j\gamma$ is
\begin{align} \label{eq_D}
\mathcal{D}_{(ij)L}&=\mathcal{D}_{(ij)L}^{\nu WW}+\mathcal{D}_{(ij)L}^{N_aVV}+ \mathcal{D}_{(ij)L}^{\nu H_1H_1} +\mathcal{D}_{(ij)L}^{N_aH_2H_2}, \crn
\mathcal{D}_{(ij)R}&=\mathcal{D}_{(ij)R}^{\nu WW}+\mathcal{D}_{(ij)R}^{N_aVV}+ \mathcal{D}_{(ij)R}^{\nu H_1H_1} +\mathcal{D}_{(ij)R}^{N_aH_2H_2}.
\end{align}

With ordinary charged leptons, we have $m_{e_i}\gg m_{e_j},i>j$ leads to $\left| \mathcal{D}_{(ji)R}\right| \gg \left| \mathcal{D}_{(ji)L}\right| $, so we usually ignore $\mathcal{D}_{(ji)L}$ in Eq.(\ref{eq_Gaebaga}) when examining $\mathrm{Br}(e_i\rightarrow e_j\gamma)$. We give the notations corresponding to the contributions to $e_i\rightarrow e_j\gamma$ decays as follows
\bea \label{componentBr}
\mathrm{Br}^{\nu}(e_i\rightarrow e_j\gamma)&&\simeq \frac{48\pi^2}{ G_F^2} \left|\sum_a \left(\mathcal{D}^{\nu_a WW}_{(ij)R} +\mathcal{D}^{\nu_a H_1H_1}_{(ji)R}\right)\right|^2 \mathrm{Br}(e_i\rightarrow e_j\overline{\nu_j}\nu_i),\crn
\mathrm{Br}^{N}(e_i\rightarrow e_j\gamma)&&\simeq \frac{48\pi^2}{ G_F^2}\left|\sum_a\left(\mathcal{D}^{N_a VV}_{(ij)R} +\mathcal{D}^{N_a H_2H_2}_{(ji)R}\right)\right|^2 \mathrm{Br}(e_i\rightarrow e_j\overline{\nu_j}\nu_i),\crn
\mathrm{Br}^{\nu W}(e_i\rightarrow e_j\gamma)&&\simeq \frac{48\pi^2}{ G_F^2}  \left|\sum_a \left(\mathcal{D}^{\nu_a WW}_{(ij)R}\right)\right|^2 \mathrm{Br}(e_i\rightarrow e_j\overline{\nu_j}\nu_i).
\eea
The third contributor ($\mathrm{Br}^{\nu W}$) consists of only the same particles as appears in the standard model. We will investigate the contributions of these components to the $\mathrm{Br}^{Total}(e_i\rightarrow e_j\gamma)$ in the numerical calculation.
\subsection{Analytic formulas for contributions to $h^0_1 \rightarrow e_i{e_j}$ decays}
\begin{figure}[ht]
	\centering
	\begin{tabular}{cc}
		\includegraphics[width=14.0cm]{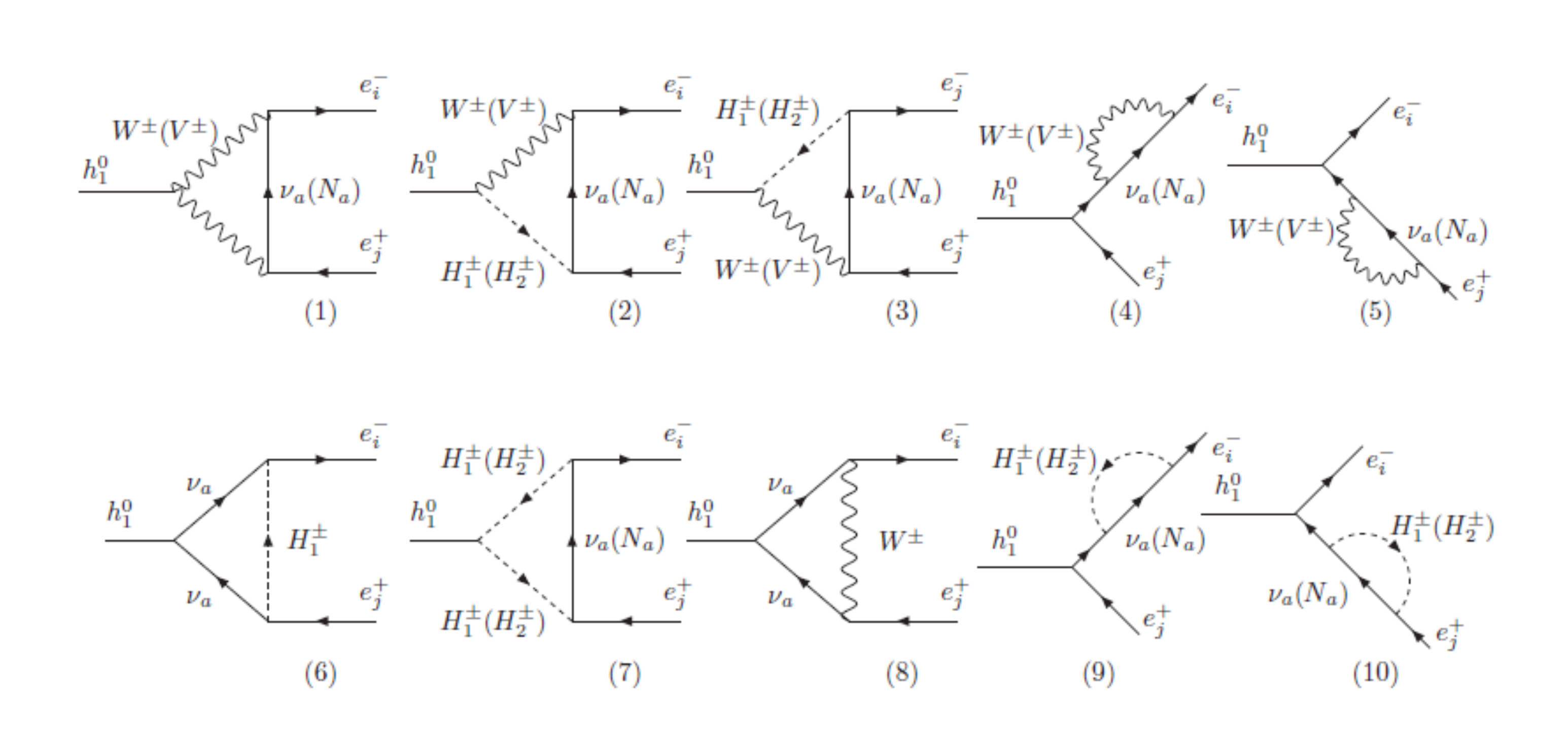} 
	\end{tabular}%
	\caption{ Feynman diagrams at one-loop order of $h_1^0 \rightarrow \mu \tau$ decays in the unitary gauge.}
	\label{fig_hmt331}
\end{figure}

For convenience when investigating the LFVHDs of the SM-like Higgs boson  $h^0_1\rightarrow e_i^{\pm}e_j^{\mp}$, we use scalar factors $\mathrm{C}_{(ij)L}$ and $\mathrm{C}_{(ij)R}$. Therefore, the effective Lagrangian of  these decays is
\bea \label{Lag_eff}
 \mathcal{L}_{\mathrm{LFVH}}^\mathrm{eff}= h^0_1 \left(\mathrm{C}_{(ij)L} \overline{e_i}P_L e_j +\mathrm{C}_{(ij)R} \overline{e_i}P_R e_j\right) + \mathrm{h.c.}
 \eea
According to the couplings listed in Tab.\ref{albga}, we obtain the one-loop Feynman diagrams contributing to these LFVHDs amplitude in the unitary gauge are shown in Fig.\ref{fig_hmt331}. Inevitablly, the scalar factors $\mathrm{C}_{(ij)L,R}$  arise from the loop contributions, we only pay attention to all corrections at one-loop order.\\

The partial width of $h^0_1\rightarrow e_i^{\pm}e_j^{\mp}$ is
\be
\Gamma (h_1^0\rightarrow e_ie_j)\equiv\Gamma (h^0_1\rightarrow e_i^{+} e_j^{-})+\Gamma (h_1^0\rightarrow e_i^{-} e_j^{+})
=  \fr{ m_{h^0_1} }{8\pi }\left(\vert \mathrm{C}_{(ij)L}\vert^2+\vert \mathrm{C}_{(ij)R}\vert^2\right). \label{LFVwidth}
\ee
We use the conditions for external momentum as: $p^2_{i,j}=m^2_{i,j}$,\, $(p_i+p_j)^2=m^2_{h^0_1}$ and $m^2_{h^0_1}\gg m^2_{i,j}$,\, this leads to branching ratio of $h^0_1\rightarrow e_i^{\pm}e_j^{\mp}$ decays can be given
\bea  Br(h_1^0\rightarrow e_ie_j)=\Gamma (h_1^0\rightarrow e_ie_j)/\Gamma^\mathrm{total}_{h^0_1}, \label{brhmt} 
\eea
where $\Gamma^\mathrm{total}_{h^0_1}\simeq 4.1\times 10^{-3}~\mathrm{GeV}$ as shown in Refs. \cite{Patrignani:2016xqp,Denner:2011mq}.

The factors corresponding to the diagrams of Fig.(\ref{fig_hmt331}) have been given in Appendix \ref{appen_loops2}. To calculate the total amplitude for LFVHD in this model, we separate into two parts, namely:  
$\mathrm{C}_{(ij)L,R}^\nu$ for the contributions of active neutrinos and $\mathrm{C}_{(ij)L,R}^{N}$ for the contributions of exotic leptons. They are
\bea  \mathrm{C}_{(ij)L,R}^\nu &=&  \sum_{a}U_{ia} U_{ja}^{*} \frac{1}{64\pi^2}\left[-g^3(c_{\alpha}s_{12}-s_\alpha c_{12})\times\mathcal{M}^{FVV}_{L,R}(m_{\nu_a},m_W)\right. \crn
&&+(-g^2(c_{\alpha}c_{12}+s_\alpha s_{12}))\times \mathcal{M}^{FVH}_{L,R} (s_{12},c_{12},v_1,v_2,m_{\nu_a},m_W,m_{H^{\pm}_1}) \crn
&&+(-g^2(c_{\alpha}c_{12}+s_\alpha s_{12}))\times \mathcal{M}^{FHV}_{L,R} (s_{12},c_{12},v_1,v_2,m_{\nu_a},m_W,m_{H^{\pm}_1}) \crn
&&+(\frac{g^3s_{\alpha}}{m_Ws_{12}})\times \mathcal{M}^{FV}_{L,R}(m_{\nu_a},m_W)\crn
&&+(\frac{-4g^3c_{\alpha}}{m_Wc_{12}})\times \mathcal{M}^{FFH}_{L,R}(s_{12},c_{12},v_1,v_2,m_{\nu_a},m_{H^{\pm}_1})\crn
&&+(-4 \lambda_{h^0H_1H_1})\times \mathcal{M}^{FHH}_{R} (s_{12},c_{12},v_1,v_2,m_{\nu_a},m_{H^{\pm}_1})\crn
&&+(\frac{-g^3c_{\alpha}}{m_Wc_{12}})\times \mathcal{M}^{VFF}_{L,R}(m_W,m_{\nu_a})\crn
&&+\left.  (\frac{4gs_{\alpha}}{m_Ws_{12}}) \times \mathcal{M}^{FH}_{L,R} (s_{12},c_{12},v_1,v_2,m_{\nu_a},m_{H^{\pm}_1})\right],  \label{nudeltaL1}\eea
and
\bea  \mathrm{C}_{(ij)L,R}^{N} &=& \sum_{a}V_{ia}^LV_{ja}^{L*}  \frac{1}{64 \pi^2}
\left[\frac{g^3s_\al c_{12}m_W}{m_V} \times \mathcal{M}^{FVV}_{L,R}(m_{N_a},m_V) \right.\crn
&&+\left(- 2 g^2s_{\alpha}c_{13}\right)\times \mathcal{M}^{FVH}_{L,R} (c_{13},s_{13},v_1,v_3,m_{N_a},m_V,m_{H^{\pm}_2}) \crn
&&+ \left(- 2 g^2s_{\alpha}c_{13}\right)\times \mathcal{M}^{FHV}_{L,R} (c_{13},s_{13},v_1,v_3,m_{N_a},m_V,m_{H^{\pm}_2}) \crn
&&+(\frac{g^3s_{\alpha}}{m_Ws_{12}})\times \mathcal{M}^{FV}_{L,R}(m_{N_a},m_V)\crn
&&+\left(- 4\lambda_{h^0_1H_2H_2}\right) \times \mathcal{M}^{FHH}_{L,R} (c_{13},s_{13},v_1,v_3,m_{N_a},m_{H^{\pm}_2})\crn
&&+\left. (\frac{4gs_{\alpha}}{m_Ws_{12}})\times \mathcal{M}^{FH}_{L,R} (c_{13},s_{13},v_1,v_3,m_{N_a},m_{H^{\pm}_2})\right].  \label{NdeltaL1}\eea
Total factor for LFVHDs process is
\bea \label{TotalAmp}
\mathrm{C}_{(ij)L,R}=\mathrm{C}_{(ij)L,R}^{\nu}+\mathrm{C}_{(ij)L,R}^{N}
\eea
In $\mathrm {C} _ {(ij) L, R}$, there are divergence terms which are implicit in PV functions ($B_0^{(n)},\,B_1^{(n)},\,n=1,2$)as shown in App.\ref{appen_loops2}. However, we can use techniques as mentioned in Refs.\cite{Hung:2021fzb,Hue:2015fbb} to separate the divergences and the finite parts in each factor.  Obviously, the divergence parts are eliminated, because their sum is zero, the finite part remaining whose contributions are shown in the following numerical investigation.
\section{Numerical results}
\label{numerical_results}
\subsection{\label{Numerical1}Setup parameters}  
We use the well-known experimental parameters \cite{Zyla:2020zbs,Patrignani:2016xqp}: 
the charged lepton masses $m_e=5\times 10^{-4}\,\mathrm{GeV}$,\,  $m_\mu=0.105\,\mathrm{GeV}$,\, $m_\tau=1.776\,\mathrm{GeV}$,\, the SM-like Higgs mass $m_{h^0_1}=125.1\,\mathrm{GeV}$,\,  the mass of the W boson $m_W=80.385\,\mathrm{GeV}$  and the gauge coupling of the $SU(2)_L$ symmetry $g \simeq 0.651$.\\
In this model we can give the relationship of the neutral gauge boson outside the standard model as $m_Z'^2=\frac{g^2v_3^2c_W^2}{3-4s_W^2}$. However, $m_Z'\geq 4.0\,\mathrm{TeV}$ is the limit given by Refs.\cite{CMS:2018ipm,ATLAS:2019erb}, resulting in $v_3\geq 10.1\, \mathrm{TeV}$. At LHC$@13\mathrm{TeV}$, we can choose $m_V=4.5[\mathrm{TeV}]$ as satisfying the above conditions. This value of $m_V$ is very suitable and will be shown in the numerical investigation below. Mixing angle between light VEVs is chosen $\frac{1}{60}\leq t_{12}\leq 3.5$ in accordance with Refs.\cite{Cepeda:2019klc,Hung:2019jue}. However, the LFVHDs in this model depend very little on the change of $t_{12}$, so we choose $t_{12}=0.5$ in the following investigations. Regarding to the $s_\delta$ parameter, this is an important parameter of THDM. In section \ref{Couplings}, we have shown that the couplings of $h^0_1$ will be similar to the standard model when $s_\delta \rightarrow 0$, combined with the condition to satisfy all THDMs then $c_\delta > 0.99$ according to the results shown in Ref.\cite{Kanemura:2018yai}, we choose $\left| s_\delta\right| < 0.14$. With this arange of $\left| s_\delta\right|$, the model under consideration also predicts the existence of large signal of $h_1^0 \rightarrow Z\ga$. This has been detailed as in recent work \cite{Hung:2019jue}. 

The absolute values of
all Yukawa and Higgs self couplings should be choose less than
$\sqrt{4\pi}$ and $4\pi$, respectively. In addition to the parameters that can impose conditions to determine the value domains, we choose the set of free parameters of this model as: $\lambda_1,\,\tilde{\lambda}_{12},\, s_\delta,\,m_{h_2^0},\,m_{N_1},\,m_{N_2}$ and $m_{H^\pm_2}$.

Therefore, the dependent parameters are given follows.
\begin{align}
\label{eq_la122}
\lambda_{2}&=\lambda_1t^4_{12} +\frac{ \left[c^2_{\delta} (1-t^2_{12}) -t_{12} s_{2\delta}\right] g^2m^2_{h^0_1} +\left[s^2_{\delta}(1-t^2_{12}) +s_{2\delta} t_{12}\right]g^2m^2_{h^0_2}}{8 c^2_{12}m_W^2} , \crn 
\lambda_{12}&=-2\lambda_1t^2_{12}  + \frac{\left( s_{2\delta} +2t_{12}c^2_{\delta} \right) g^2m^2_{h^0_1} +\left(  2s^2_{\delta}t_{12}-s_{2\delta} \right)g^2m^2_{h^0_2}}{8 s_{12}c_{12} m_W^2},\crn
\lambda_{23}&=\frac{s^2_{12}}{v_3^2} \left[  m^2_{h^0_1} +m^2_{h^0_2} -\frac{8m_W^2}{g^2}\left( \lambda_1s_{12}^2 +\lambda_2c_{12}^2  \right)\right],
\end{align}
and $\lambda_{13}$ was given by using the invariance trace of the squared mass matrices in Eq.(\ref{eq-mixingh0}) in App.\ref{appen_HGB} as,
\begin{equation}\label{eq_la13} 
		\lambda_{13}=\frac{c^2_{12}}{v_3^2} \left[  m^2_{h^0_1} +m^2_{h^0_2} -\frac{8m_W^2}{g^2}\left( \lambda_1s_{12}^2 +\lambda_2c_{12}^2  \right)\right] . 
\end{equation}

Regarding to the parameters of active neutrinos we use the recent results of experiment as shown in Refs.~\cite{Patrignani:2016xqp,Tanabashi:2018oca,Zyla:2020zbs}: 
$\Delta m_{21}^2=7.55 \times 10^{-5} \mathrm{eV}^2$,\, $\Delta m_{31}^2=2.424 \times 10^{-3} \mathrm{eV}^2$,\, $\sin^2 \theta^\nu_{12}=0.32$,\, $\sin^2 \theta^\nu_{23}=0.547$,\, $\sin^2 \theta^\nu_{13}=0.0216$.

The mixing matrix of active neutrinos is derived from the $U^{MNPS}$ when we ignore a very small deviation \cite{Ibarra:2010xw}. That way, one gives $U \equiv U^L=U (\theta_{12}^\nu,\theta_{13}^\nu,\theta_{23}^\nu)$ and $U^\dagger=U^\dagger (\theta_{12}^\nu,\theta_{13}^\nu,\theta_{23}^\nu)$, with $\theta_{ij}^\nu$ are mixing angles of active neutrinos, the parameterized form of $U$ matrix is
\bea U (\theta_{12},\theta_{13},\theta_{23})&=&\left(
\begin{array}{ccc}
	1 & 0 & 0 \\
	0 & \cos\theta_{23} & \sin\theta_{23} \\
	0 & -\sin\theta_{23} & \cos\theta_{23} \\
\end{array}
\right)
\left(
\begin{array}{ccc}
	\cos\theta_{13} & 0 & \sin\theta_{13} \\
	0 & 1 & 0 \\
	-\sin\theta_{13} & 0 & \cos\theta_{13} \\
\end{array}
\right)\crn
&\times&
\left(
\begin{array}{ccc}
	\cos\theta_{12} & \sin\theta_{12} & 0 \\
	-\sin\theta_{12} & \cos\theta_{12} & 0 \\
	0 & 0 & 1 \\
\end{array}
\right). \label{mixingpar}\eea


Exotic leptons are also mixed in a common way based on Eq.~(\ref{mixingpar}), by choosing $V^L \equiv U^L(\theta_{12}^N,\theta_{13}^N,\theta_{23}^N)$,  with $\theta_{ij}^N$ are mixing angles of exotic leptons. The  parameterization of $V^L$ is chosen so that the LFV decays can be obtained large signals. According to that criterion, we can give some cases corresponding to large mixing angle of exotic leptons and there are following interesting cases: $V^L \equiv U^L(\frac{\pi}{4},\frac{\pi}{4},\frac{\pi}{4})$,\hs$V^L \equiv U^L(\frac{\pi}{4},\frac{\pi}{4},-\frac{\pi}{4})$ and $V^L \equiv U^L(\frac{\pi}{4},0,0)$. The other cases only change minus signs in the total amplitudes without changing the final result of branching ratios of the LFVHD process.

\subsection{\label{Numerical2} Numerical results of cLFV}  
Analytical results for the components of $e_i \rightarrow e_j\ga$ have been given at Eq.(\ref{componentBr}), using them we give the parameter space domains of this model satisfying the experimental limits of $e_i \rightarrow e_j\ga$ decays. 

\begin{figure}[ht]
	\centering
	\begin{tabular}{cc}
		\includegraphics[width=7.0cm]{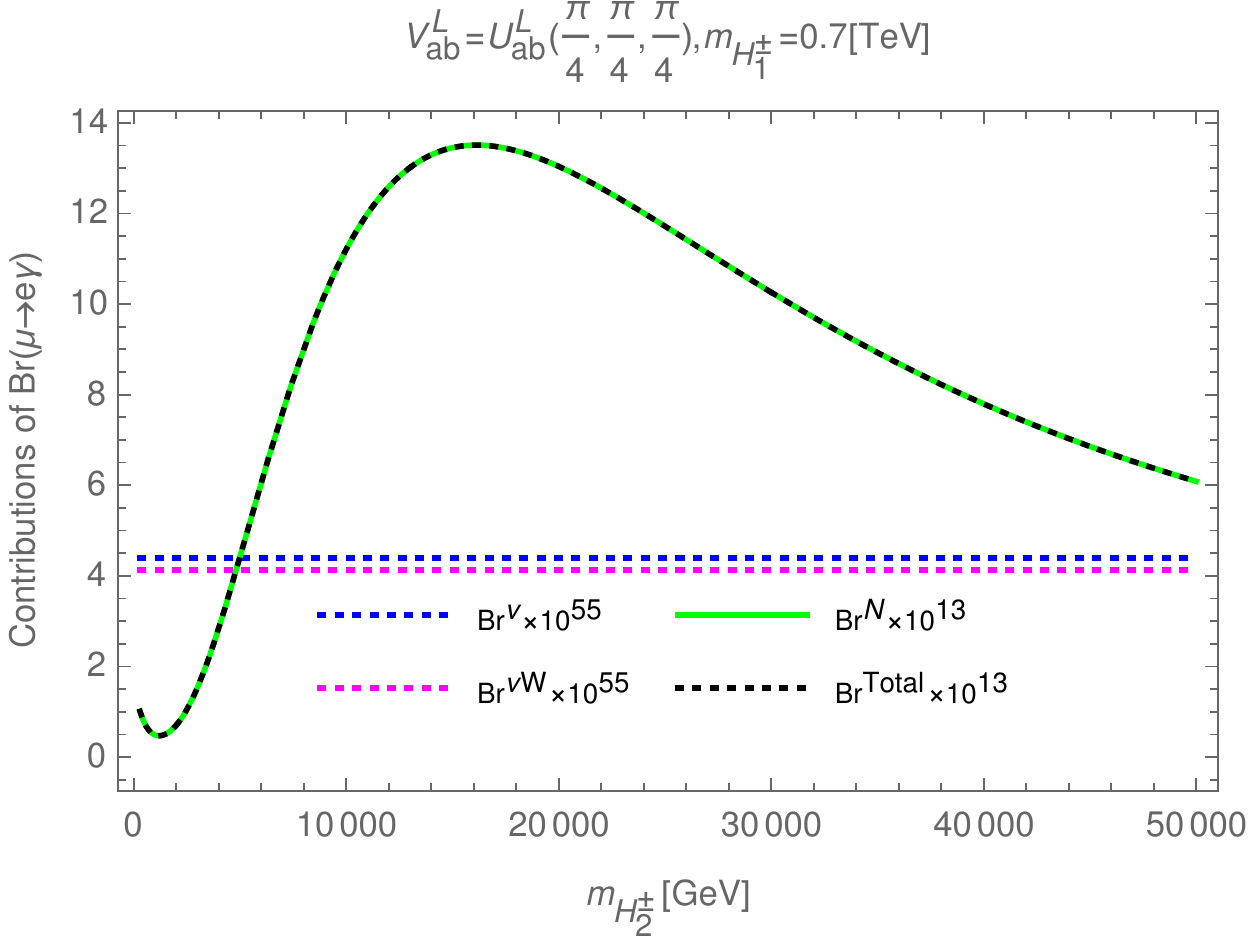}& 
		\includegraphics[width=7.0cm]{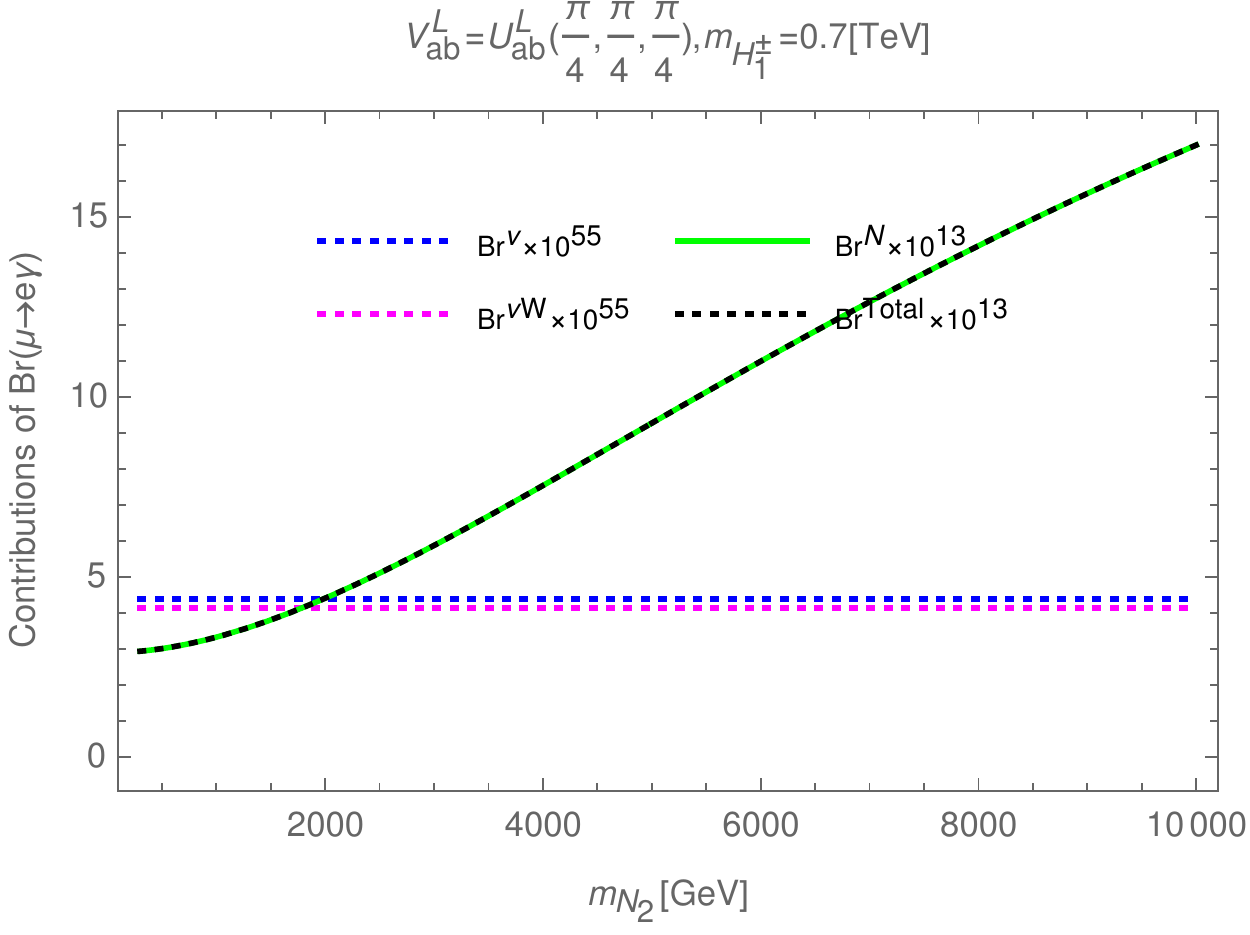}
	\end{tabular}%
	\caption{ The contributions to $\mu \rightarrow e\gamma$ decay in the case of $V^L_{ab} = U^L_{ab}(\pi/4,\pi/4,\pi/4)$ depend on $m_{H^\pm_2}$ (left panel) or $m_{N_2}$ (right panel).}
	\label{fig_Comp}
\end{figure}

Among cLFV decays, $\mu \rightarrow e\ga$  has the strictest experimental limit, so in the regions of parameter space where $\mu \rightarrow e\ga$ satisfies the experimental limits, the $\tau \rightarrow e\ga$ and $\tau \rightarrow \mu \ga$ decays also satisfy. This result has been shown in the same studies as mentioned in Refs.\cite{Hue:2017lak,Hung:2021fzb}. To avoid unnecessary investigations, we only introduce parameter regions satisfying the experimental conditions of $\mu \rightarrow e\ga$ in different mixing cases of exotic leptons. These domains are ensured matching for the remaining two cLFV decays.\\
Without loss of generality, we can perform ahead with the $V^L_{ab} = U^L_{ab}(\pi/4,\pi/4,\pi/4)$ case. Then, the components of $\mu \rightarrow e\gamma$ decay are given as Fig.(\ref{fig_Comp}).

As a result, $\mathrm{Br}^\nu$ and $\mathrm{Br}^{\nu W}$ give a very small contribution compared to $\mathrm{Br}^{Total}$, while the $\mathrm{Br}^N$ of the exotic leptons is very close to the main contribution. So, we can ignore small contributions in the later calculations. In particular, with the choice of the largest mixing parameters of the exotic leptons, significant signals for $\mu \rightarrow e\gamma$ decay are around $m_{H^\pm_2}<8.0~ \mathrm{TeV}$ (left panel) or around $m_{N_2}<2.0~ \mathrm{TeV}$ (right panel).\\

We also mentioned about the anomalous magnetic moments of electron
	and muon $a_{e,\mu}=(g_{e,\mu}-2)/2$, which are of interest now. But they are closely related to the decays of charged leptons. From Eqs.(\ref{eq_Gaebaga},\ref{eq_D}), we can write,
\bea
a_{e_i}=-\frac{4m^2_{e_i}}{e}\mathrm{Re}\left[\mathcal{D}_{(ii)R} \right]= -\frac{4m^2_{e_i}}{e}\left( \mathrm{Re}\left[\mathcal{D}^\nu_{(ii)R} \right]+\mathrm{Re}\left[\mathcal{D}^N_{(ii)R} \right]\right), \label{amu}
\eea 
where $\mathcal{D}^\nu_{(ij)R} =\mathcal{D}^{\nu_a WW}_{(ij)R}+\mathcal{D}^{\nu_a H_1H_1}_{(ij)R},\,\mathcal{D}^{N}_{(ij)R}=\mathcal{D}^{N_a VV}_{(ij)R}+\mathcal{D}^{N_a H_2H_2}_{(ij)R}$, using the results in App.\ref{appen_loops1}, we have:
\bea
\mathcal{D}^\nu_{(ij)R} \sim \sum_a m^2_{\nu_a}U_{ia}U_{aj}^\dagger, \hs \mathcal{D}^{N}_{(ij)R} \sim \sum_a m^2_{N_a} V_{ia}V_{aj}^*.
\eea
According to the results obtained in Fig.\ref{fig_Comp}, the contribution of active neutrinos is very small compared to that of neutral leptons. Therefore, the contributions to the anomalous magnetic moments of muon and cLFV are
\bea 
\mathcal{D}_{(21)R} \simeq \mathcal{D}^N_{(21)R} \sim \sum_a m^2_{N_a}V_{2a}V_{a1}^*, \hs a_\mu \simeq -\frac{4m^2_{\mu}}{e}\mathrm{Re}\left[\mathcal{D}^{N}_{(22)R}\right]  \sim \sum_am^2_{N_a}V_{2a}V_{a2}^*.
\eea
 With the form of the matrix $V^L_{ab}$ chosen as Eq.(\ref{mixingpar}), then $\mathcal{D}_{(21)R}$ and $\mathcal{D}_{(22)R}$ are always of the same order. Furthermore, in the limit $\mathrm{Br}(\mu \rightarrow e\gamma)\leq 4.2 \times 10^{-13}$ , the condition $|\mathcal{D}_{(21)R}|\leq \mathcal{O}(10^{-13})$ is needed. Hence, $|\mathcal{D}_{(22)R}|\leq \mathcal{O}(10^{-13})$ \cite{Hue:2021xap}, resulting in $|\Delta a_\mu|\leq \mathcal{O}(10^{-13})$ . This is a very small signal compared to the current experimental limit ($\mathcal{O}(10^{-9})$ ), so the signal of $|\Delta a_\mu|$ is negligible in the regions of the parameter space that we choose to investigate the cLFV decays.

\begin{figure}[ht]
	\centering
	\begin{tabular}{cc}
		\includegraphics[width=7.0cm]{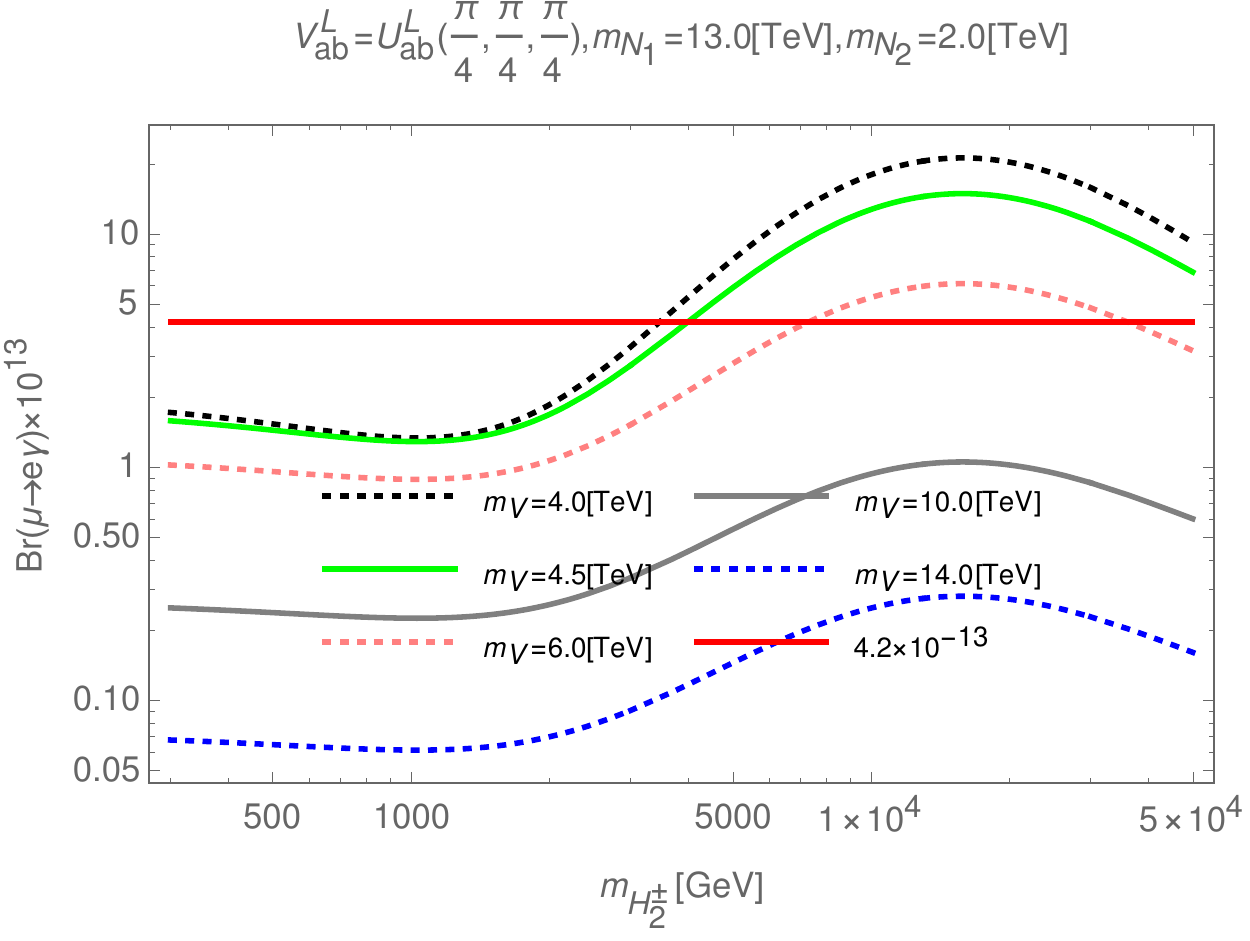}&
		\includegraphics[width=7.0cm]{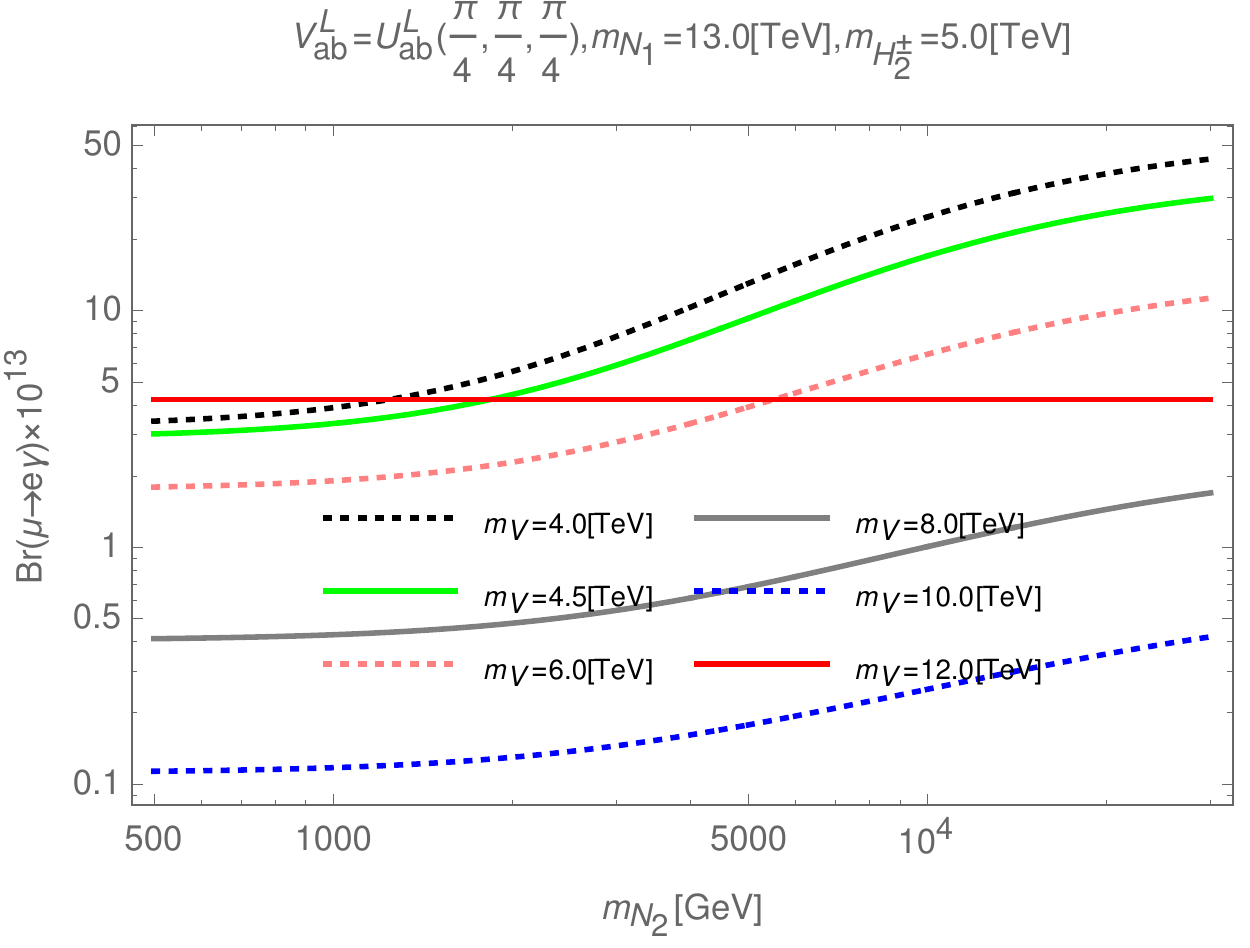}  
	\end{tabular}%
	\caption{Dependences of $\mathrm{Br}(\mu \rightarrow e\gamma)$ on $m_{V}$ and $m_{H^\pm_2}$ (left) or $m_{N_2}$ (right)  in the case of $V^L_{ab} = U^L_{ab}(\pi/4,\pi/4,\pi/4)$.}
	\label{fig_mV}
\end{figure}

Based on Refs.\cite{CMS:2018ipm,ATLAS:2019erb}, in this model ($\beta =-\frac{1}{\sqrt{3}}$) we have the limit of heavy VEV of $v_3 \geq 10.1 ~ \mathrm{TeV}$ resulting in $m_V\geq 3.8 ~ \mathrm{TeV}$. Correspondingly, at the expected energy scale of the LHC, $v_3 \sim 13 ~ \mathrm{TeV}$, then $m_V\sim 4.5 ~ \mathrm{TeV}$. To select the appropriate region of $m_V$, we fix $m_{N_1}=13~\mathrm{TeV}$, then the dependences of $\mathrm{Br}(\mu \rightarrow e\gamma)$ on $m_V$ and $m_{H^\pm_2}$ or $m_{N_2}$ are given as Fig.(\ref{fig_mV}).

As the result in Fig.(\ref{fig_mV}), the larger the values of $m_V$, the better the experimental limits of $\mathrm{Br}(\mu \rightarrow e\gamma)$ are satisfied. However, this appears undesirable that the value of $\mathrm{Br}(\mu \rightarrow e\gamma)$ is small which makes  it difficult to detect experimentally. We found the best fit when choosing $m_V=4.5~\mathrm{TeV}$ to perform the  numerical investigation of lepton-flavor-violating decays.  

\begin{figure}[ht]
	\centering
	\begin{tabular}{cc}
		\includegraphics[width=7.0cm]{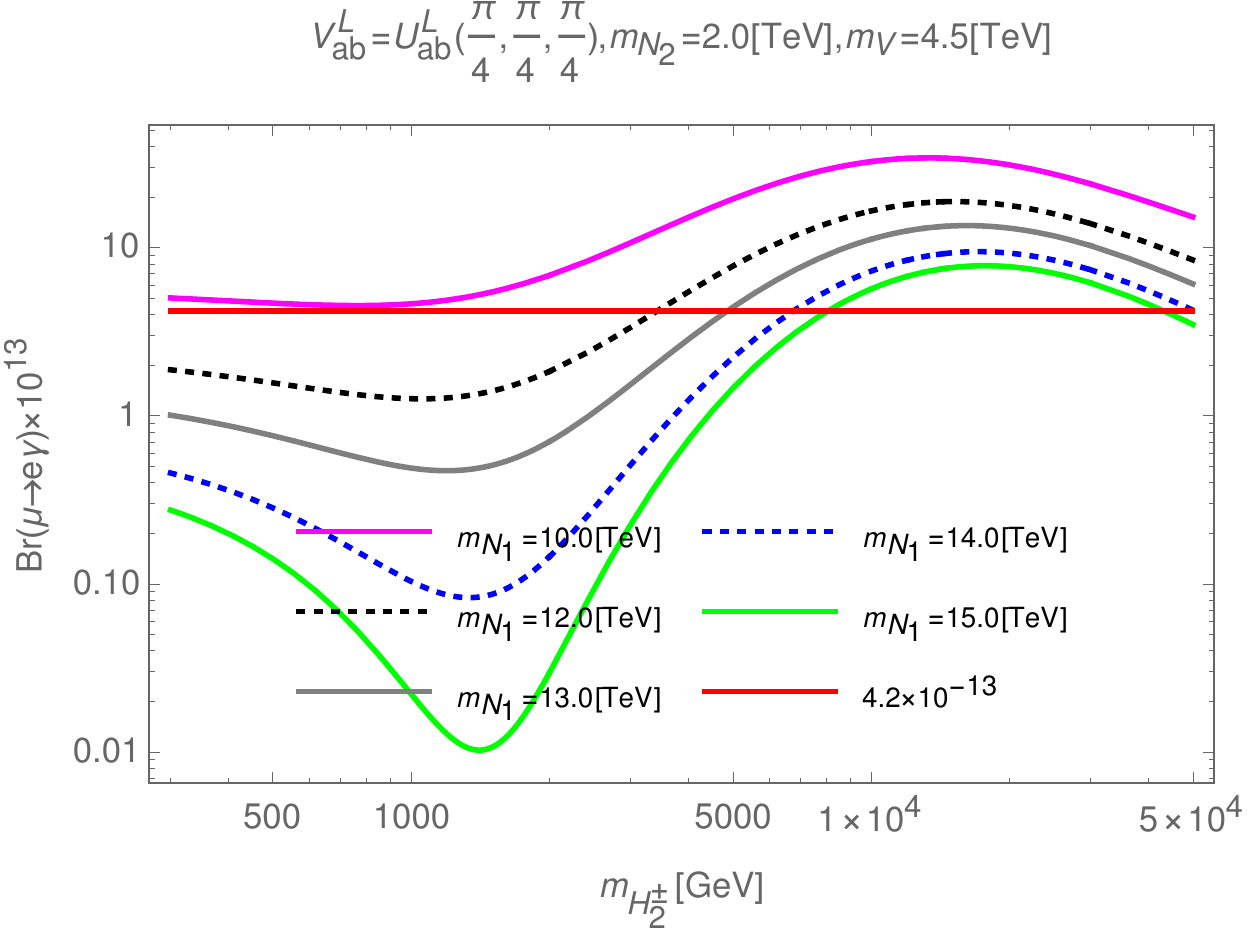}& \includegraphics[width=7.0cm]{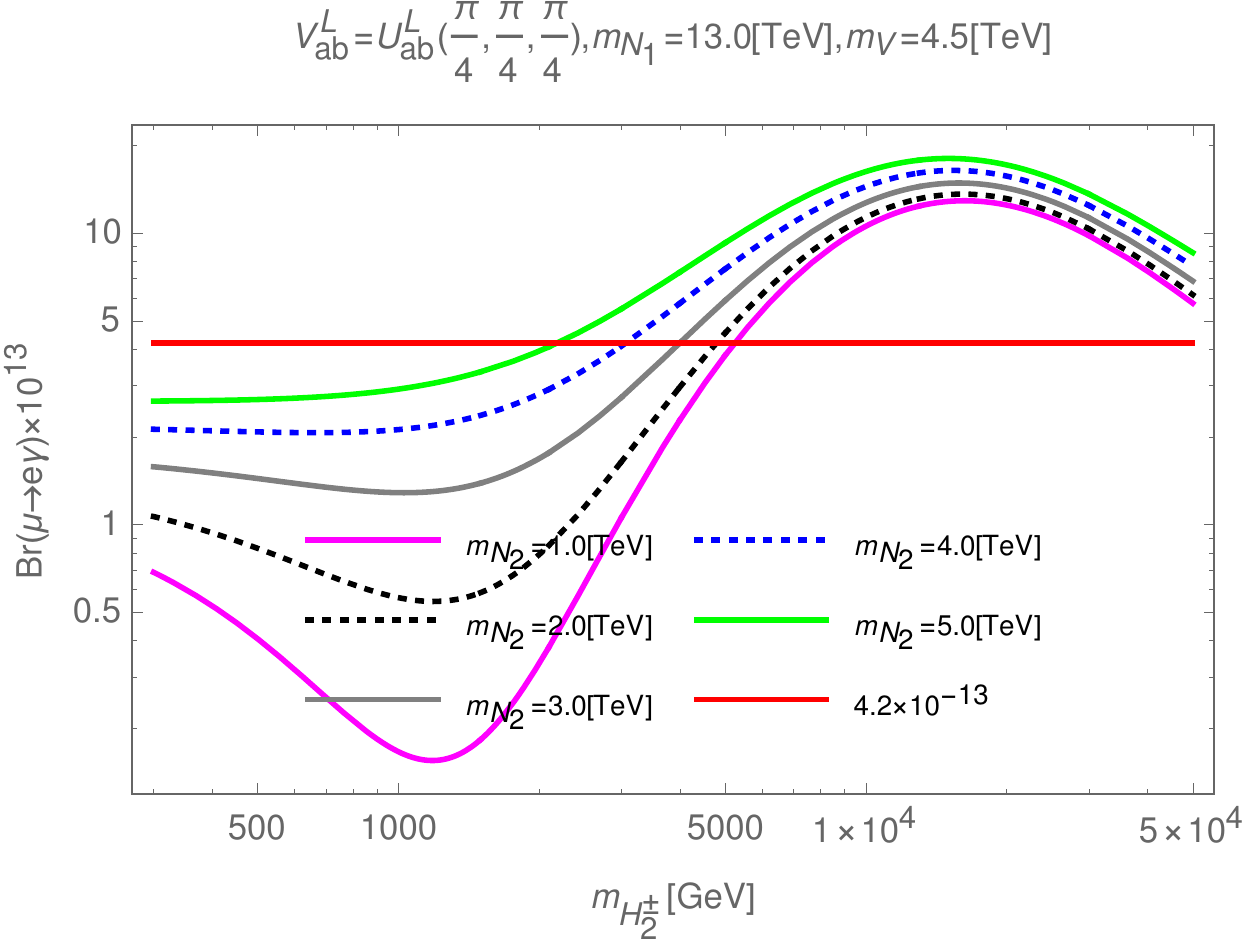}\\
		\includegraphics[width=7.0cm]{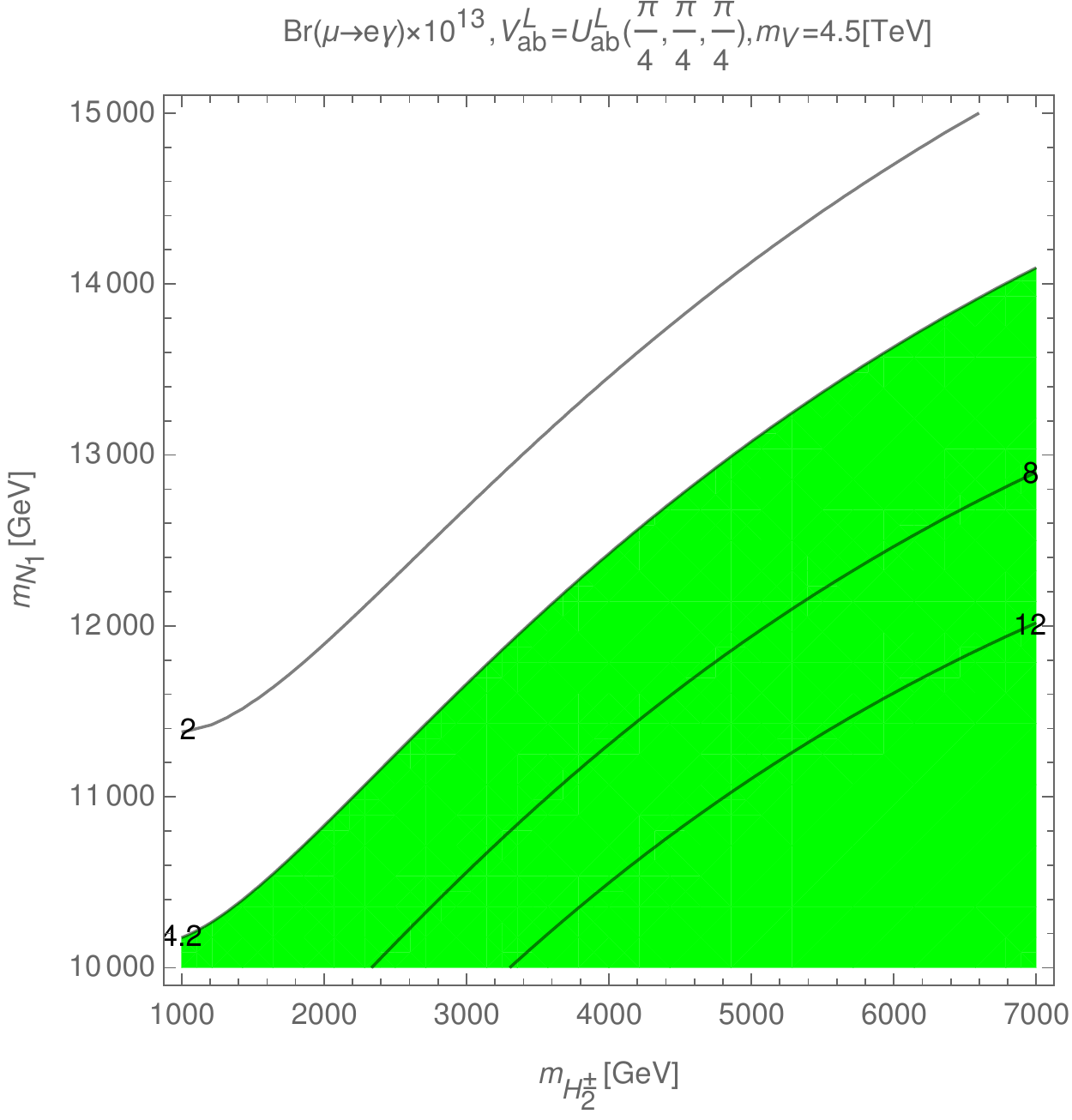}& \includegraphics[width=7.0cm]{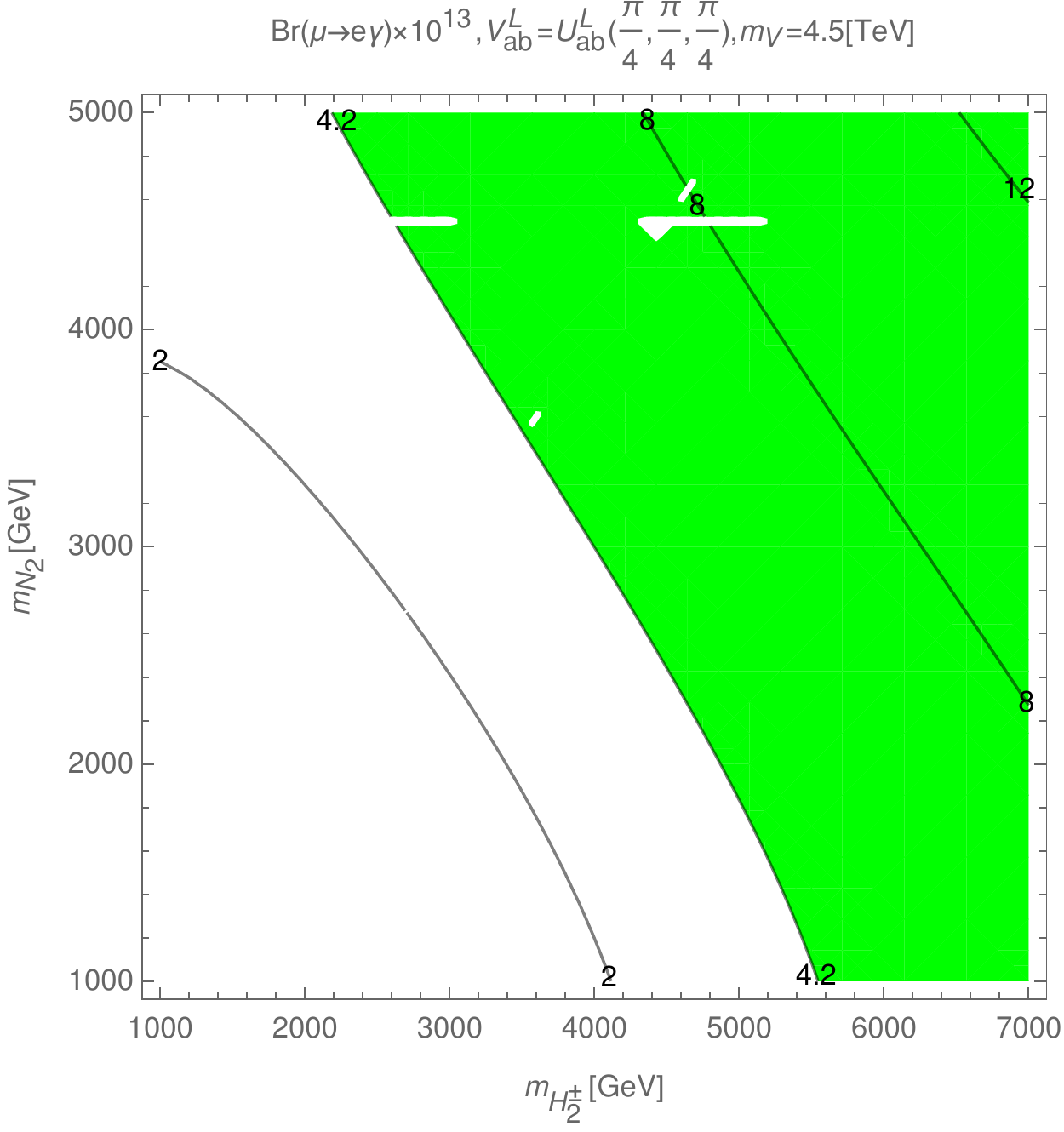}
	\end{tabular}%
	\caption{ Dependence of $\mathrm{Br}(\mu \rightarrow e\gamma)$ on $m_{H^\pm_2}$ (first row) and contour plots of $\mathrm{Br}(\mu \rightarrow e\gamma)$ (second row) as functions of $m_{H^\pm_2}$ and $m_{N_1}$ (left panel) or $m_{H^\pm_2}$ and $m_{N_2}$ (right panel) in the case of $V^L_{ab} = U^L_{ab}(\pi/4,\pi/4,\pi/4)$.}
	\label{fig_mNamNb}
\end{figure}

The results of the numerical survey show that $\mathrm{Br}(\mu \rightarrow e\gamma)$ depends very little on the change of $t_{12}$. Therefore, to ensure the limit $\frac{1}{60}\leq t_{12}\leq 3.5$, we always choose the fixed value $t_{12}=0.5$. Combined with the fixed selection of $m_V=4.5~\mathrm{TeV}$, the dependence of $\mathrm{Br}(\mu \rightarrow e\gamma)$ on $m_{H^\pm_2}$ and the masses of exotic leptons is given as Fig.(\ref{fig_mNamNb}).

In Fig.(\ref{fig_mNamNb}), we consider the dependence of $\mathrm{Br}(\mu \rightarrow e\gamma)$ on $m_{H^\pm_2}$ and $m_{N_1}$ or $m_{N_2}$. As the result in first row, we show the parameter space region to $\mathrm{Br}(\mu \rightarrow e\gamma)<4.2\times 10^{-13}$ in the domain $1.0 ~\mathrm{TeV} \leq m_{H^\pm_2}\leq 7.0~\mathrm{TeV} $ with two cases: i) $m_{N_1}=2.0~\mathrm{TeV}$ and $m_{N_2}$ is around $13.0 ~\mathrm{TeV}$ (left) or ii) $m_{N_2}=2.0~\mathrm{TeV}$ and $m_{N_1}$ is around $13.0 ~\mathrm{TeV}$ (right). In each of these parameter regions, the value curves of $\mathrm{Br}(\mu \rightarrow e\gamma)$ decrease as $m_{N_1}$ increases (left) or increase as $m_{N_2}$ increases (right). Therefore, regarding the contributions of $m_{N_1}$ and $m_{N_2}$ to $\mathrm{Br}(\mu \rightarrow e\gamma)$ in this case, we can give in short form as: $m_{N_2}$ has an increasing effect, whereas $m_{N_1}$ has a decreasing effect. This property is also true for the other two decays, $\tau \rightarrow e\gamma$ and $\tau \rightarrow \mu\gamma$. The combination of these properties leads to the existence of regions of parameter space that satisfy the experimental limits of $e_i \rightarrow e_j\gamma$ decays when one exotic lepton has a mass about $2.0 ~\mathrm{TeV}$ and another is about $13.0 ~\mathrm{TeV}$. These significant space regions are shown to correspond to the colorless part as shown in second row of Fig.(\ref{fig_mNamNb}).

\begin{figure}[ht]
	\centering
	\begin{tabular}{cc}
		\includegraphics[width=7.0cm]{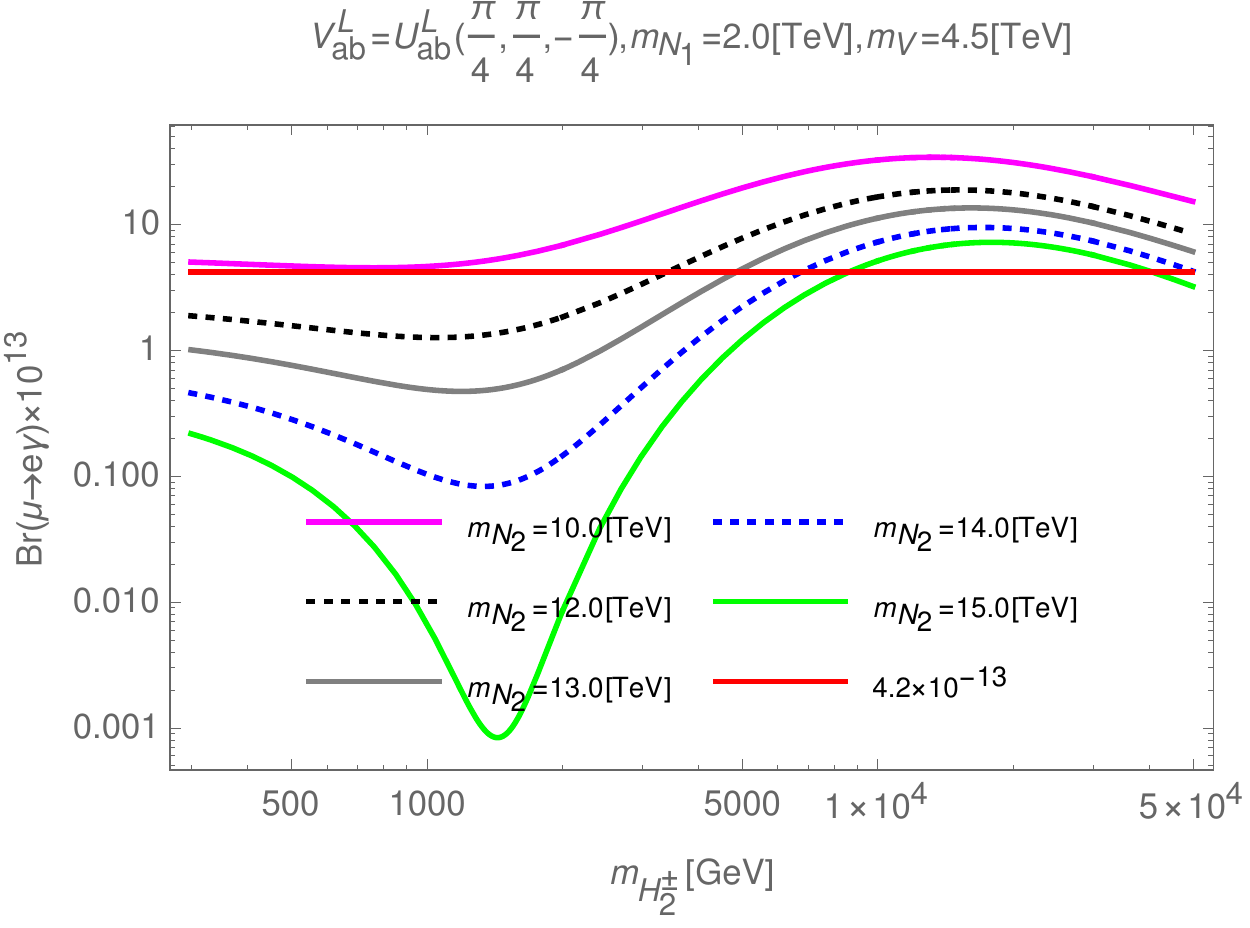}& \includegraphics[width=7.0cm]{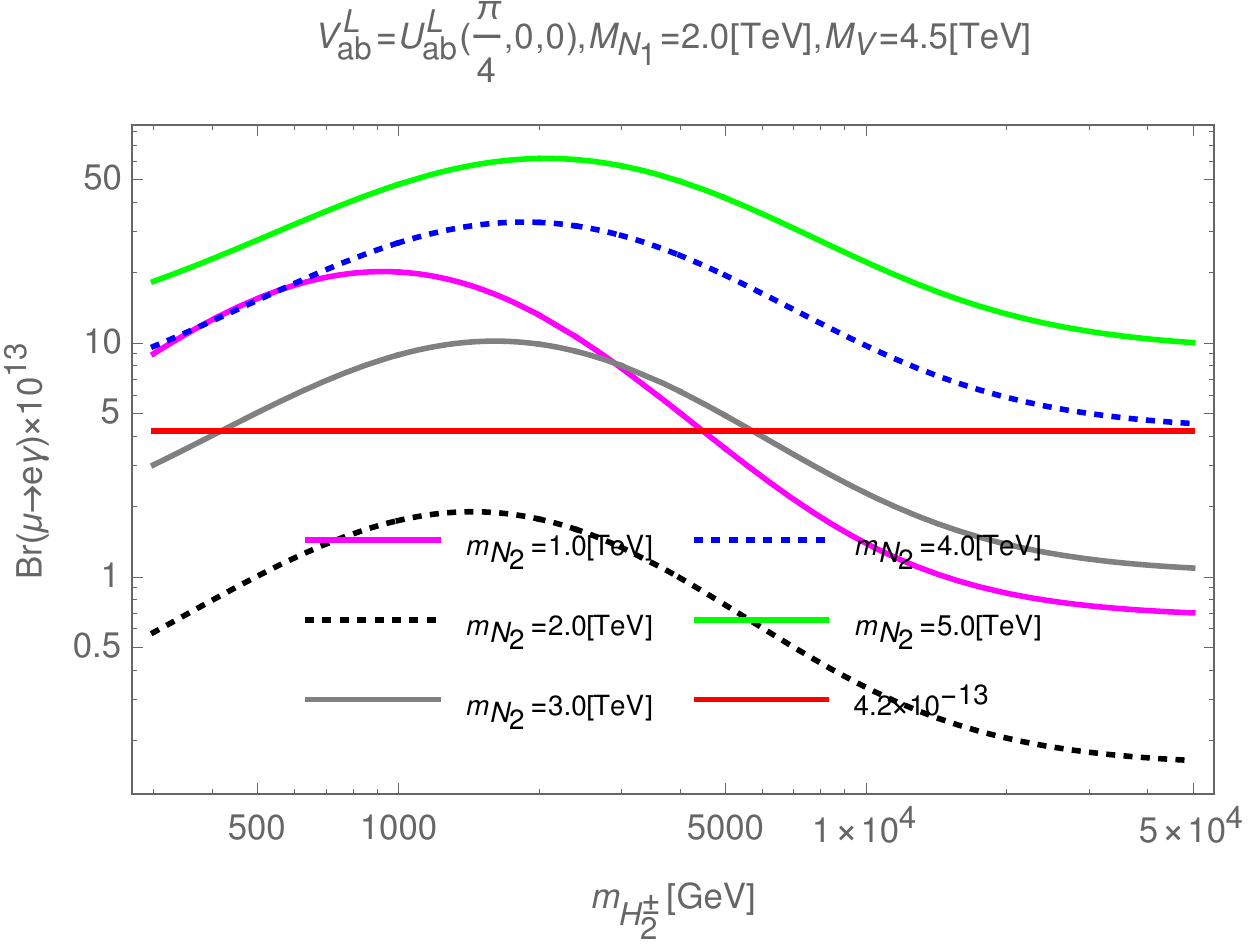}\\
		\includegraphics[width=7.0cm]{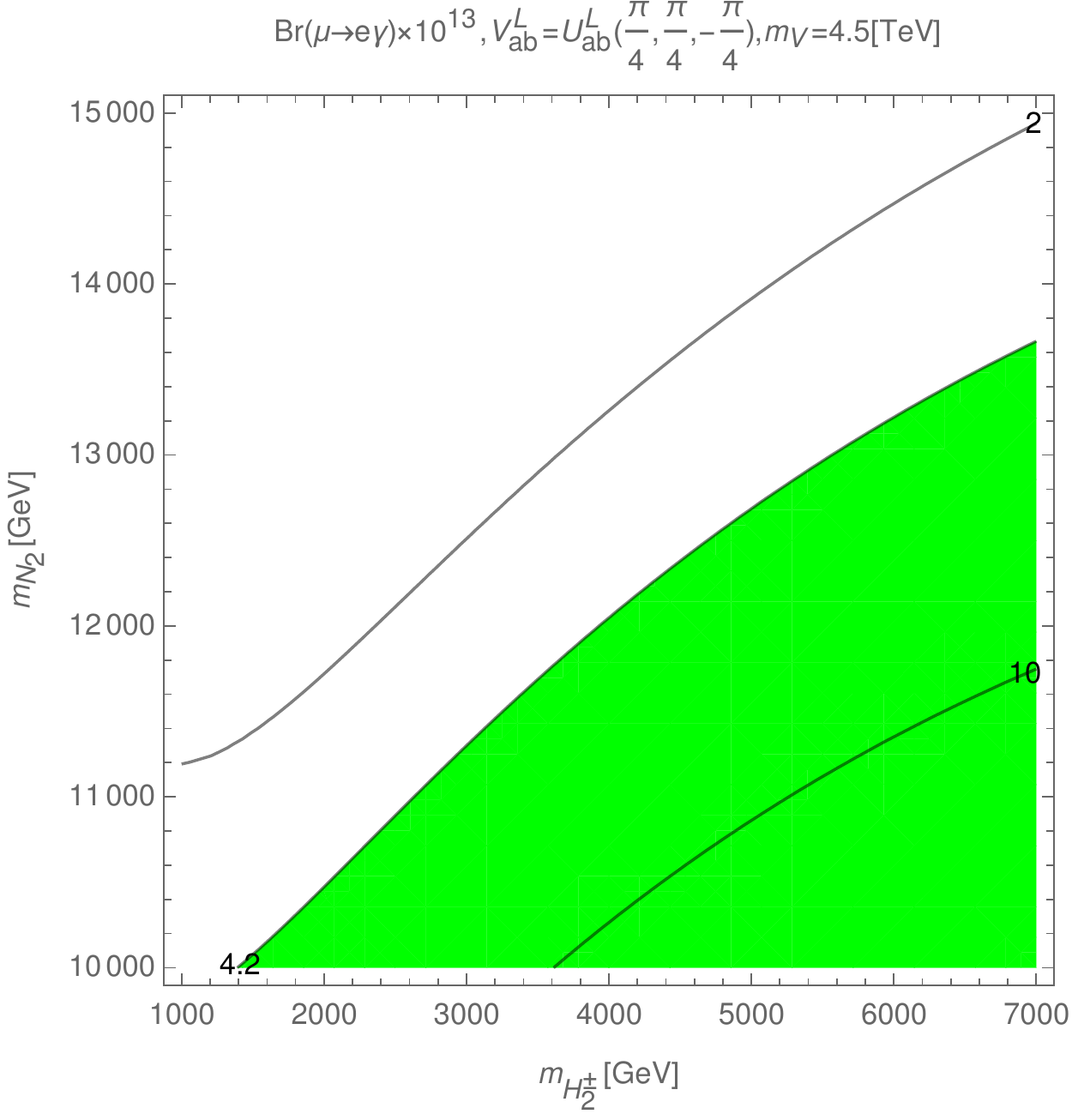}& \includegraphics[width=7.0cm]{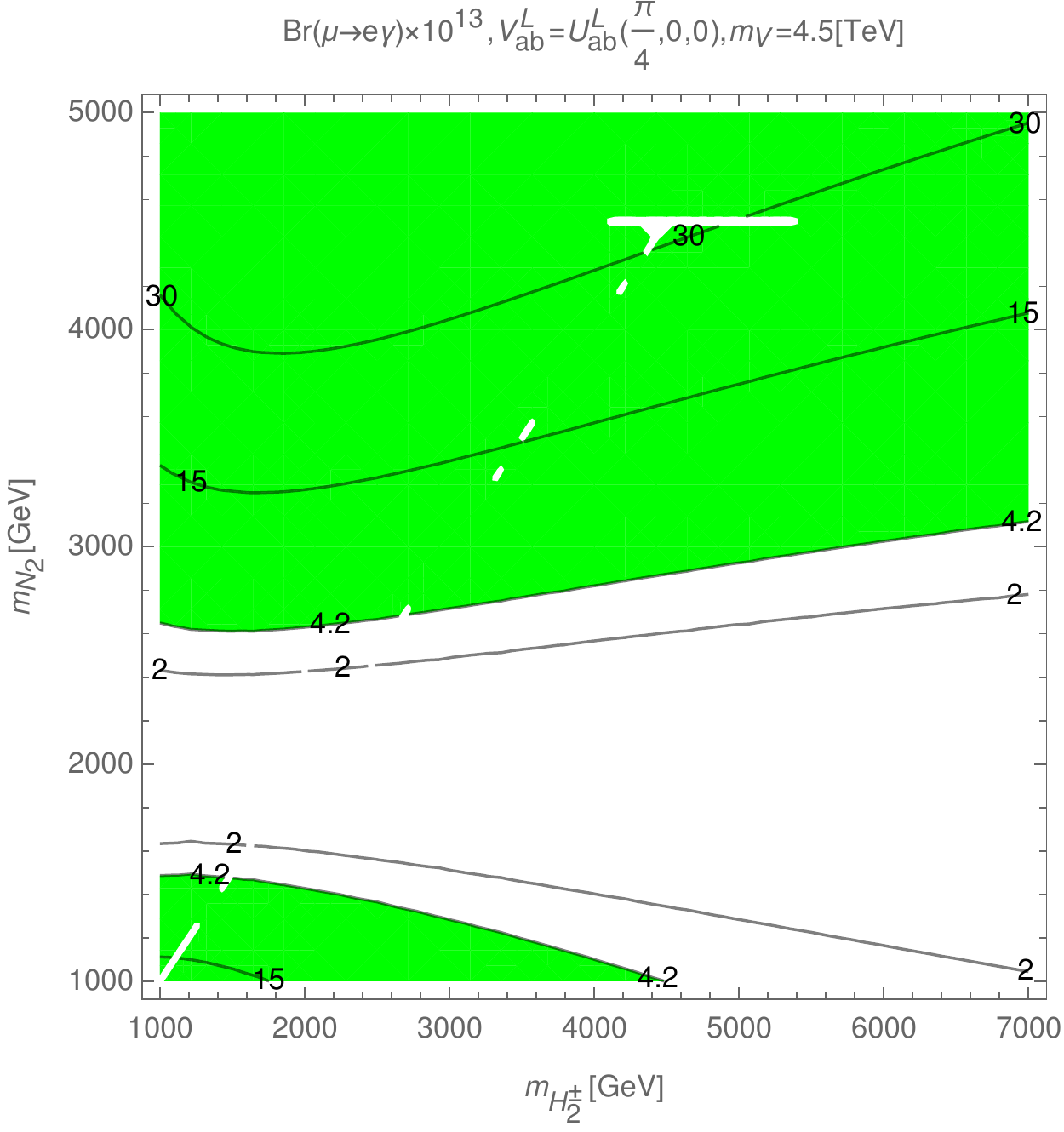}
	\end{tabular}%
	\caption{ Plots of $\mathrm{Br}(\mu \rightarrow e\gamma)$ depending on $m_{H^\pm_2}$ (first row) and contour plots of $\mathrm{Br}(\mu \rightarrow e\gamma)$ as functions of $m_{H^\pm_2}$ and $m_{N_2}$ (second row) in the case of $V^L_{ab} = U^L_{ab}(\pi/4,\pi/4,-\pi/4)$ (left panel) or in the case of $V^L_{ab} = U^L_{ab}(\pi/4,0,0)$ (right panel).}
	\label{fig_mVv}
\end{figure}

In exactly the same way, we can give results of the remaining typical cases of exotic leptons mixing as Fig.(\ref{fig_mVv}).

In both $V^L_{ab} = U^L_{ab}(\pi/4,\pi/4,-\pi/4)$ and $V^L_{ab} = U^L_{ab}(\pi/4,0,0)$ cases, we choose $m_{N_1}=2.0 ~\mathrm{TeV}$, then the allowed domains of the $e_i \rightarrow e_j\gamma$ decays are shown in the second row of Fig.(\ref{fig_mVv}). We found that the $V^L_{ab} = U^L_{ab}(\pi/4,\pi/4,-\pi/4)$ and $V^L_{ab} = U^L_{ab}(\pi/4,\pi/4,\pi/4)$ cases give nearly the same results when the role of $m_{N_1}$ and $m_{N_2}$ were swapped (see at left panels in second row in Fig.(\ref{fig_mNamNb}) and Fig.(\ref{fig_mVv})). This is also a typical feature of this model, so the numerical investigation below are mainly performed according to the dependence on $m_{H^\pm_2}$ and $m_{N_2}$.

\subsection{\label{Numerical3} Numerical results of LFVHD} 
The three decays of LFVHDs have an experimental upper limit as given in Eq.(\ref{hmt-limmit}). We can investigate these decays in the regions of the parameter space that satisfy $e_i \rightarrow e_j\gamma$ decays. From Eq.(\ref{TotalAmp}) and App.\ref{appen_loops2}, we realize that $C_{(ij)L}\sim m_j,\,C_{(ij)R}\sim m_i$ combined with $m_\tau \gg m_\mu \gg m_e$, so $\mathrm{Br}(h^0_1 \rightarrow \mu\tau)$ can receive the largest signal among the LFVHDs in this model. Therefore, we focus on finding the large signal of $h^0_1 \rightarrow \mu\tau$ decay in the following surveys.

\begin{figure}[ht]
	\centering
	\begin{tabular}{cc}
		\includegraphics[width=7.0cm]{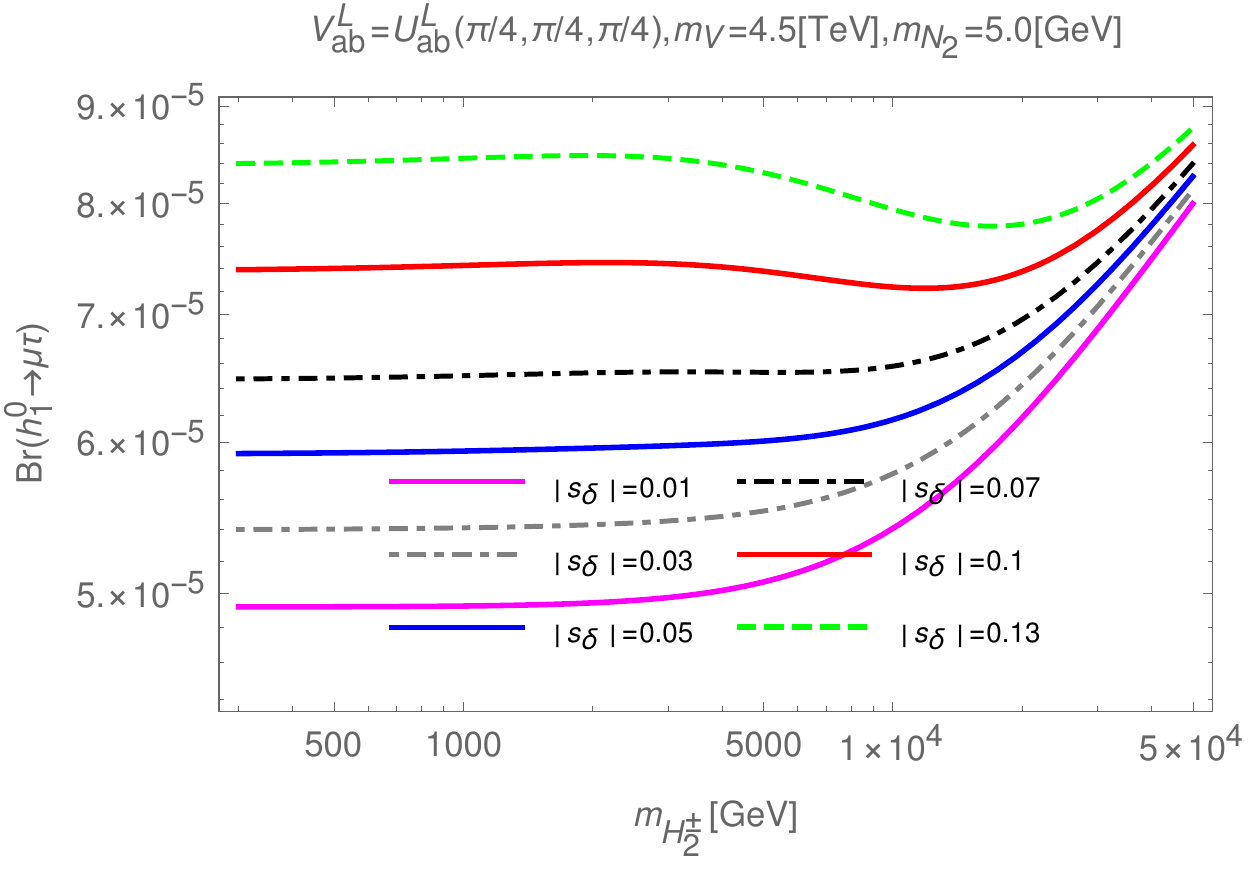}& 
		\includegraphics[width=7.0cm]{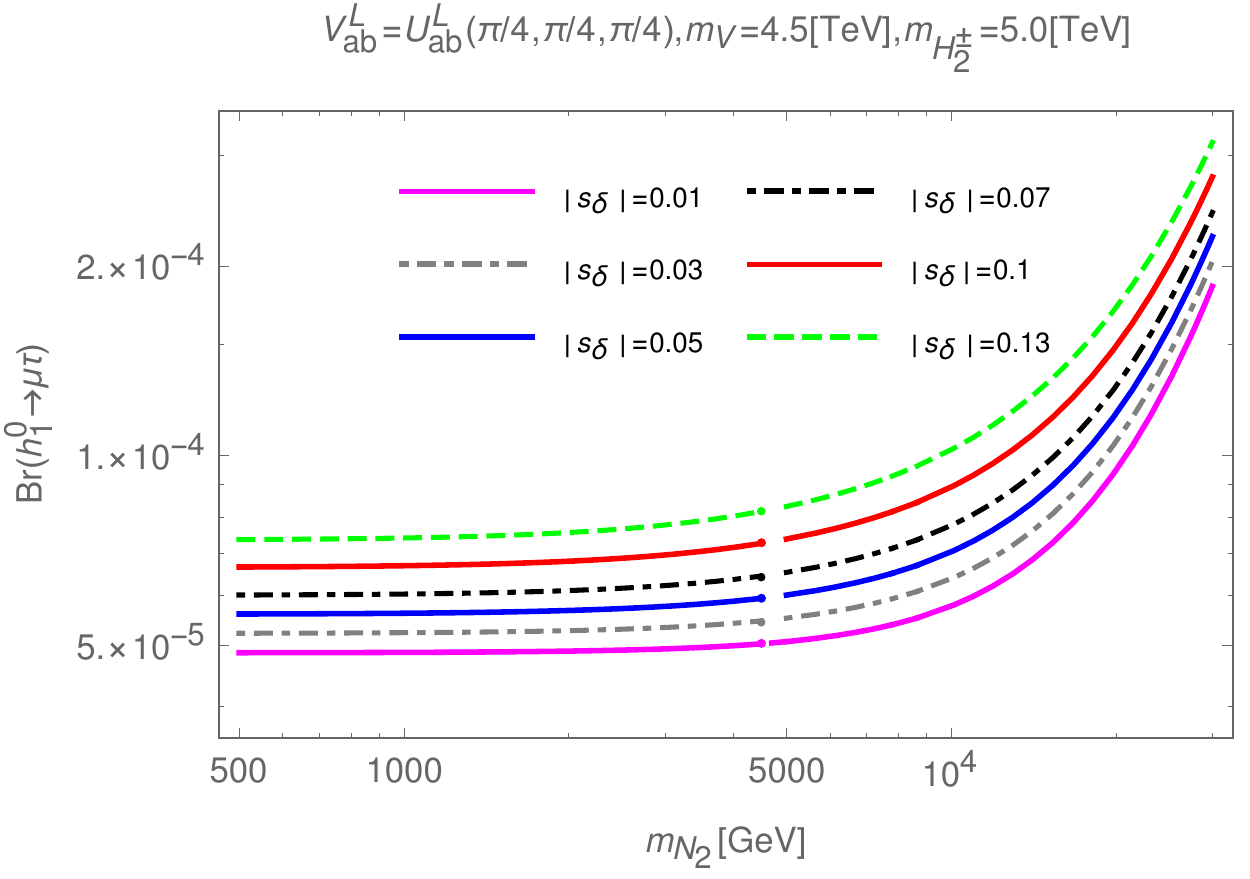}
	\end{tabular}%
	\caption{ Plots $\mathrm{Br}(h^0_1 \rightarrow \mu\tau)$ depending on $m_{H^\pm_2}$ (left) or $m_{N_2}$ (right) in case of $V^L_{ab} = U^L_{ab}(\pi/4,\pi/4,\pi/4)$ .}
	\label{fig_sd1}
\end{figure}

We use Eq.(\ref{brhmt}) to investigate the dependence of $\mathrm{Br}(h^0_1 \rightarrow \mu\tau)$ on $s_\delta$ in case of $V^L_{ab} = U^L_{ab}(\pi/4,\pi/4,\pi/4)$ , ($s_\delta$-specific parameter for THDM), the results are given as shown in Fig.(\ref{fig_sd1}). Obviously, $\mathrm{Br}(h^0_1 \rightarrow \mu\tau)$ increases proportionally to the absolute value of $s_\delta$. Therefore, in the limited range $0<\left|  s_\delta \right| <0.14$, $\mathrm{Br}(h^0_1 \rightarrow \mu\tau)$ can give the largest signal when $\left|  s_\delta \right| \rightarrow 0.14$. 

\begin{figure}[ht]
	\centering
	\begin{tabular}{cc}
		\includegraphics[width=7.0cm]{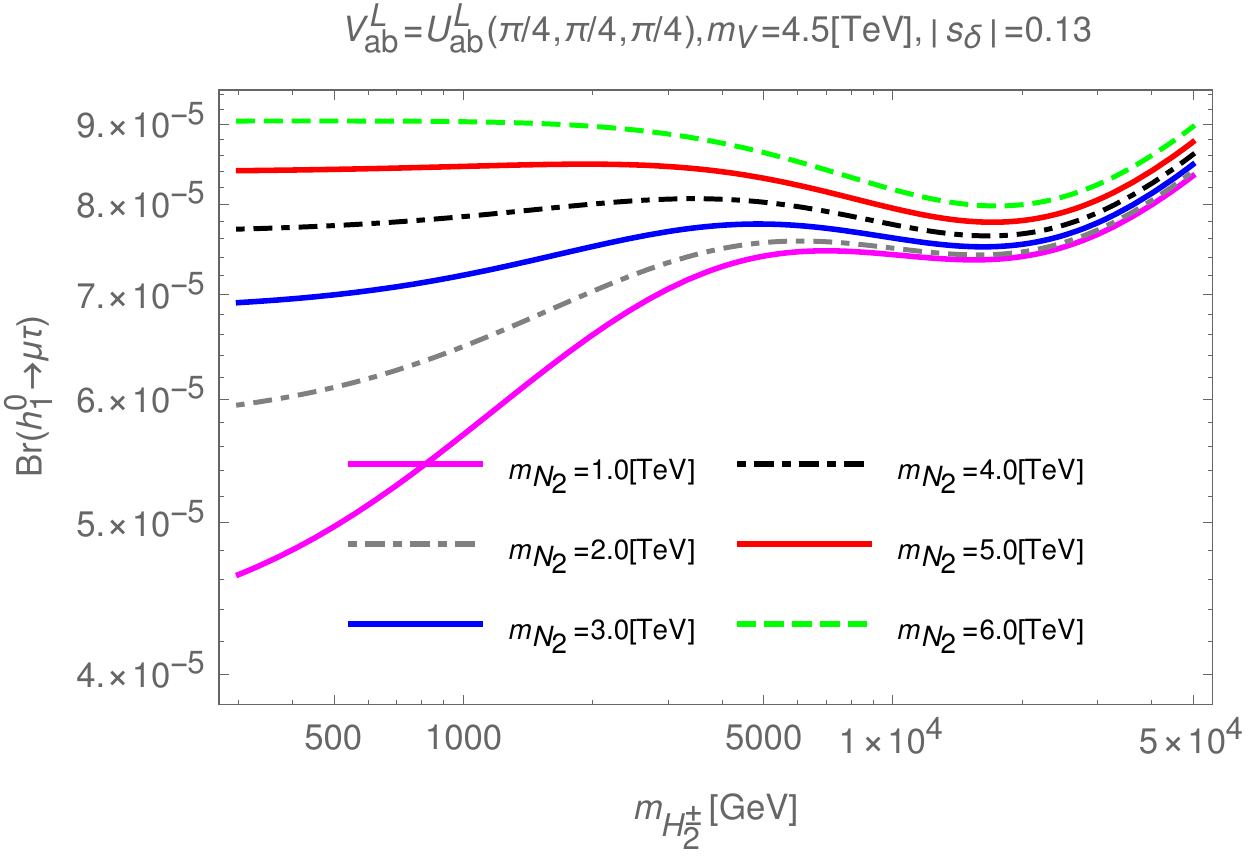}& \includegraphics[width=7.0cm]{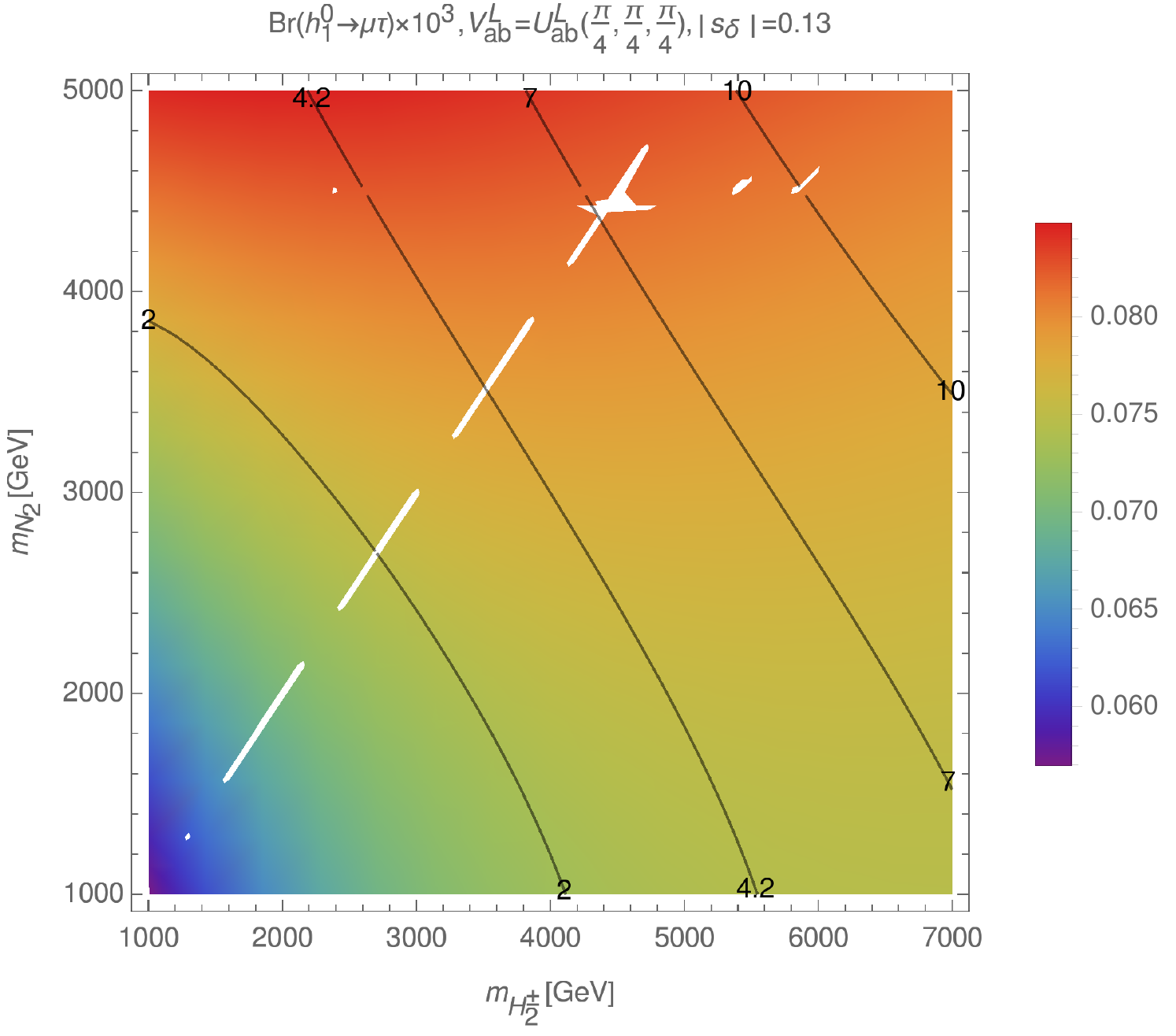}
	\end{tabular}%
	\caption{Plots $\mathrm{Br}(h^0_1 \rightarrow \mu\tau)$ as function of $m_{H^\pm_2}$ in case of $V^L_{ab} = U^L_{ab}(\pi/4,\pi/4,\pi/4)$ (left) and density plots of $\mathrm{Br}(h^0_1 \rightarrow \mu\tau)$ as function of $m_{H^\pm_2}$ and $m_{N_2}$ (right). The black cuvers in the right panel show the constant values of $\mathrm{Br}(\mu \rightarrow e\gamma)\times 10^{13}$.}
	\label{fig_p13}
\end{figure}

In the case of $V^L_{ab} = U^L_{ab}(\pi/4,\pi/4,\pi/4)$ , to find the possible parameter space for large signal of $\mathrm{Br}(h^0_1 \rightarrow \mu\tau)$, we choose a fixed value $\left|  s_\delta \right| = 0.13$, then the change of $\mathrm{Br}(h^0_1 \rightarrow \mu\tau)$ according to $m_{H^\pm_2}$ and $m_{N_2}$ is shown as Fig.(\ref{fig_p13}).

As a result, $\mathrm{Br}(h^0_1 \rightarrow \mu\tau)$ increases with $m_{N_2}$ as the left part of Fig.(\ref{fig_p13}), however the part of the parameter space is really significant, where the experimental limits of $\mathrm{Br}(\mu \rightarrow e\gamma)$ are satisfied is shown in the interval between the curves $4.2$ in the right part of Fig.(\ref{fig_p13}). We show that the largest value that $\mathrm{Br}(h^0_1 \rightarrow \mu\tau)$ can achieve in this case is about $\mathcal{O}(10^{-5})$.

\begin{figure}[ht]
	\centering
	\begin{tabular}{cc}
		\includegraphics[width=7.0cm]{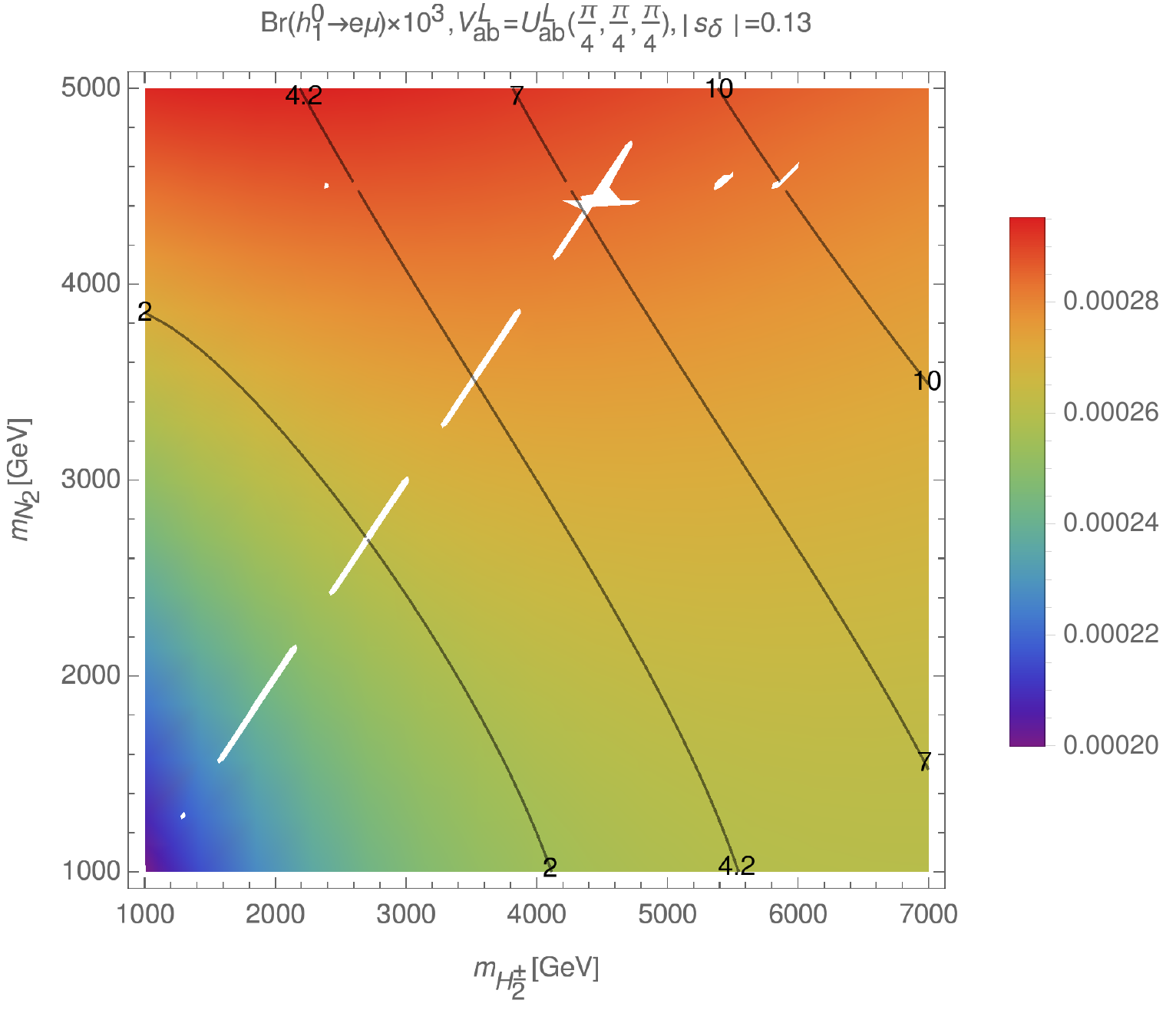}& \includegraphics[width=7.0cm]{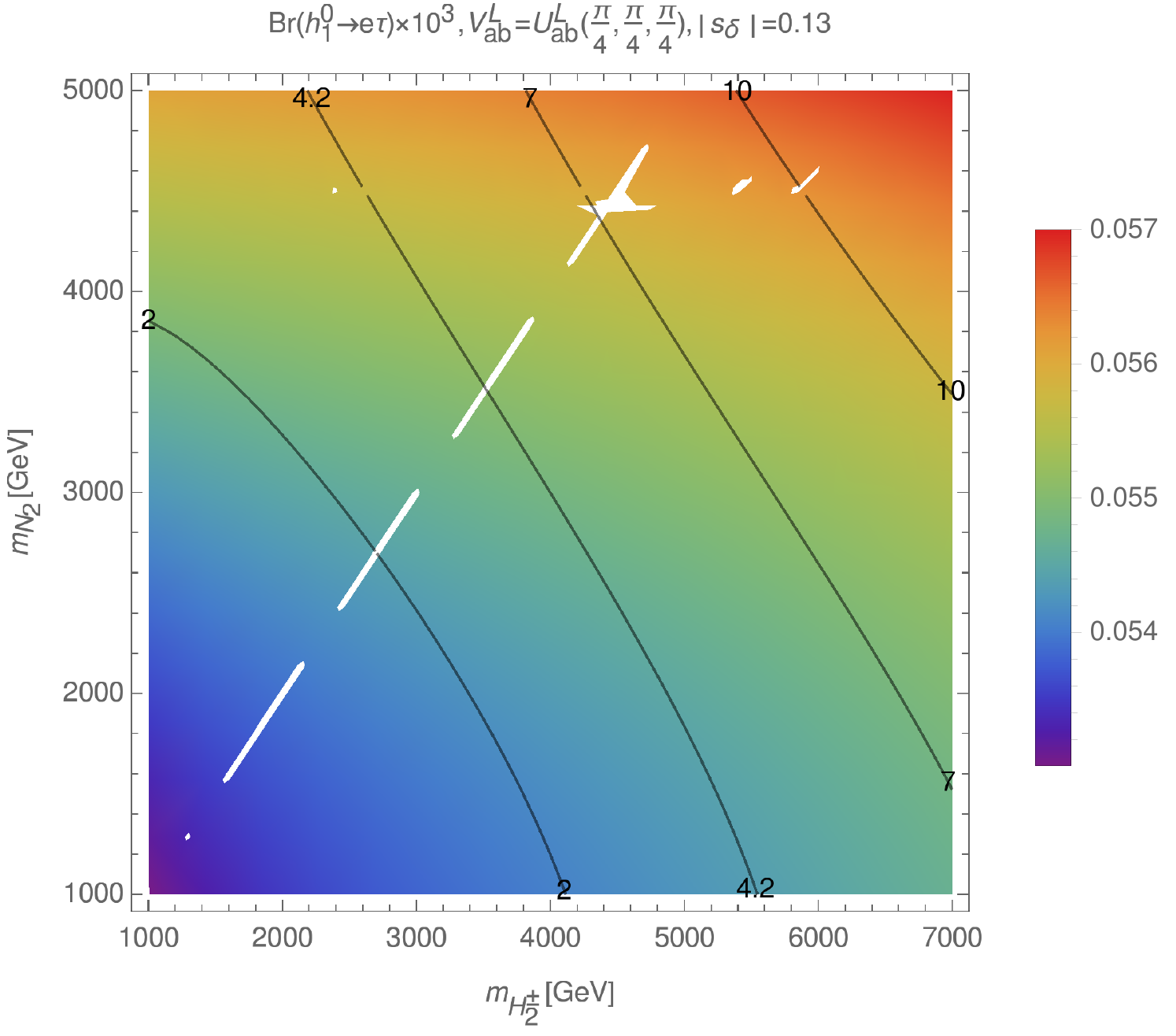}
	\end{tabular}%
	\caption{Density plots of $\mathrm{Br}(h^0_1 \rightarrow e\mu)$ (left) and  $\mathrm{Br}(h^0_1 \rightarrow e\tau)$ (right) as function of $m_{H^\pm_2}$ and $m_{N_2}$ in case of $V^L_{ab} = U^L_{ab}(\pi/4,\pi/4,\pi/4)$. The black cuvers show the constant values of $\mathrm{Br}(\mu \rightarrow e\gamma)\times 10^{13}$.}
	\label{fig_p14}
\end{figure}

In a similar way, we investigate the $\mathrm{Br}(h^0_1 \rightarrow e\mu)$ and $\mathrm{Br}(h^0_1 \rightarrow e\tau)$  in the region of the parameter space given in Fig.(\ref{fig_p13}). The results are shown in Fig.(\ref{fig_p14}).

The results obtained of $\mathrm{Br}(h^0_1 \rightarrow e\mu)$  and $\mathrm{Br}(h^0_1 \rightarrow e\tau)$  are below the upper bound of the experimental limits as mentioned in Eq.(\ref{hmt-limmit}). In addition, these values are smaller than the corresponding ones of $\mathrm{Br}(h^0_1 \rightarrow \mu \tau)$. Therefore, we are only interested in the large signal that $\mathrm{Br}(h^0_1 \rightarrow \mu \tau)$ can be achieved in the other investigation cases.

\begin{figure}[ht]
	\centering
	\begin{tabular}{cc}
		\includegraphics[width=7.0cm]{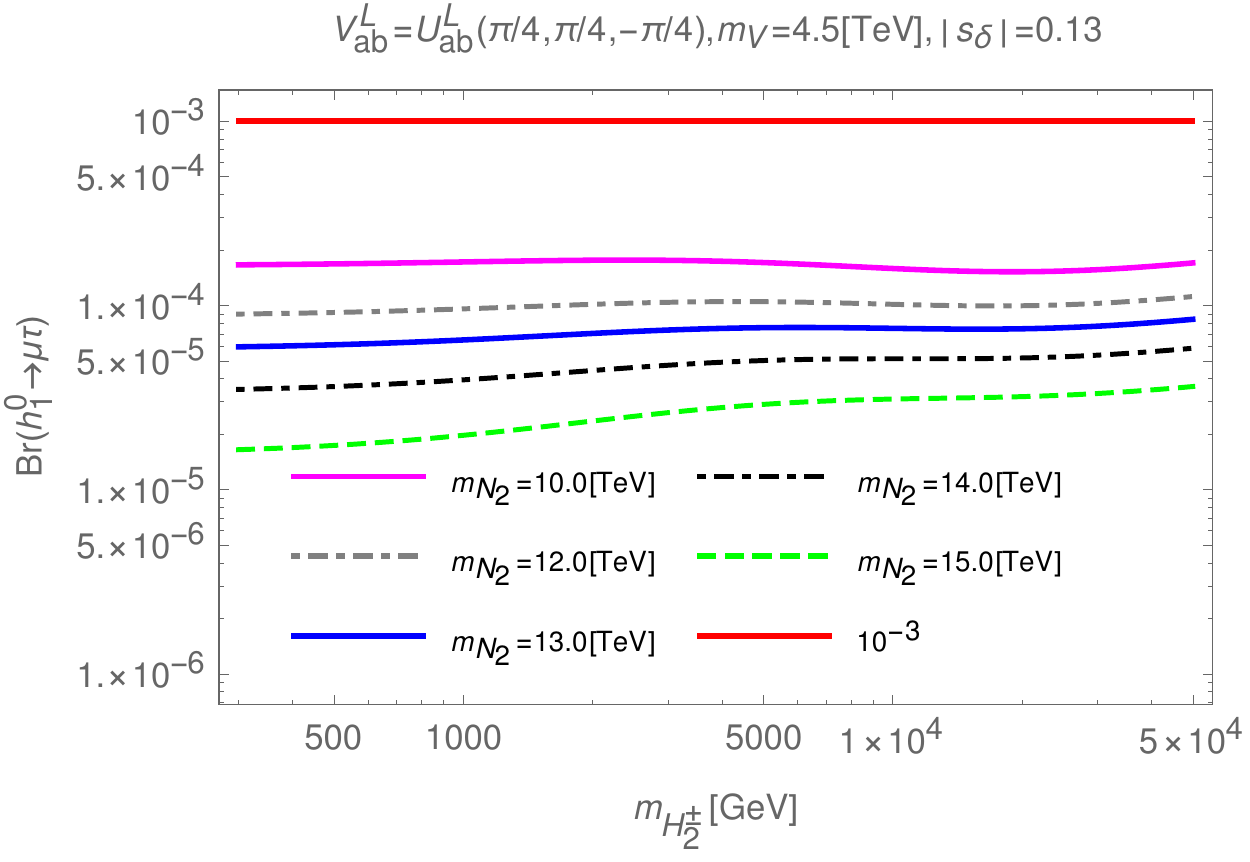}& \includegraphics[width=7.0cm]{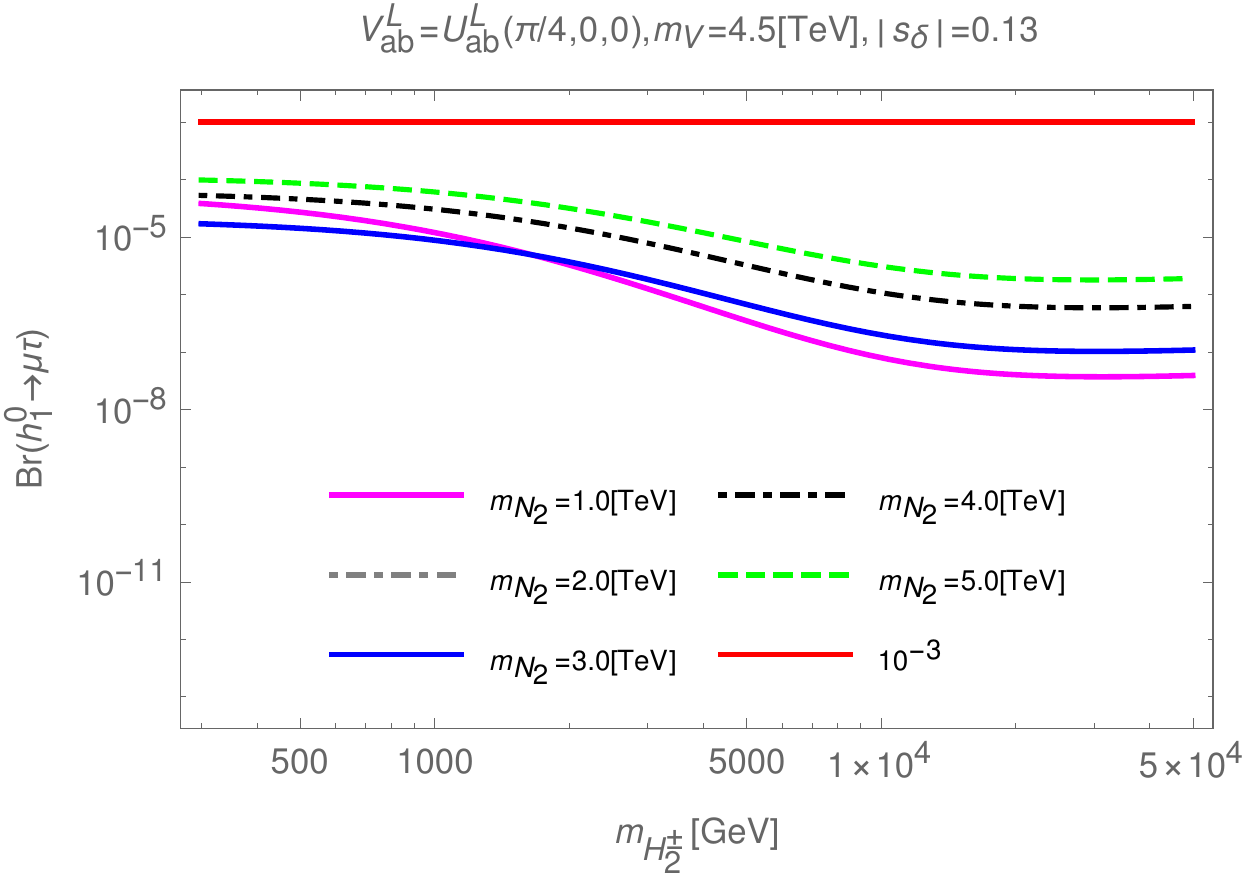}\\
		\includegraphics[width=7.0cm]{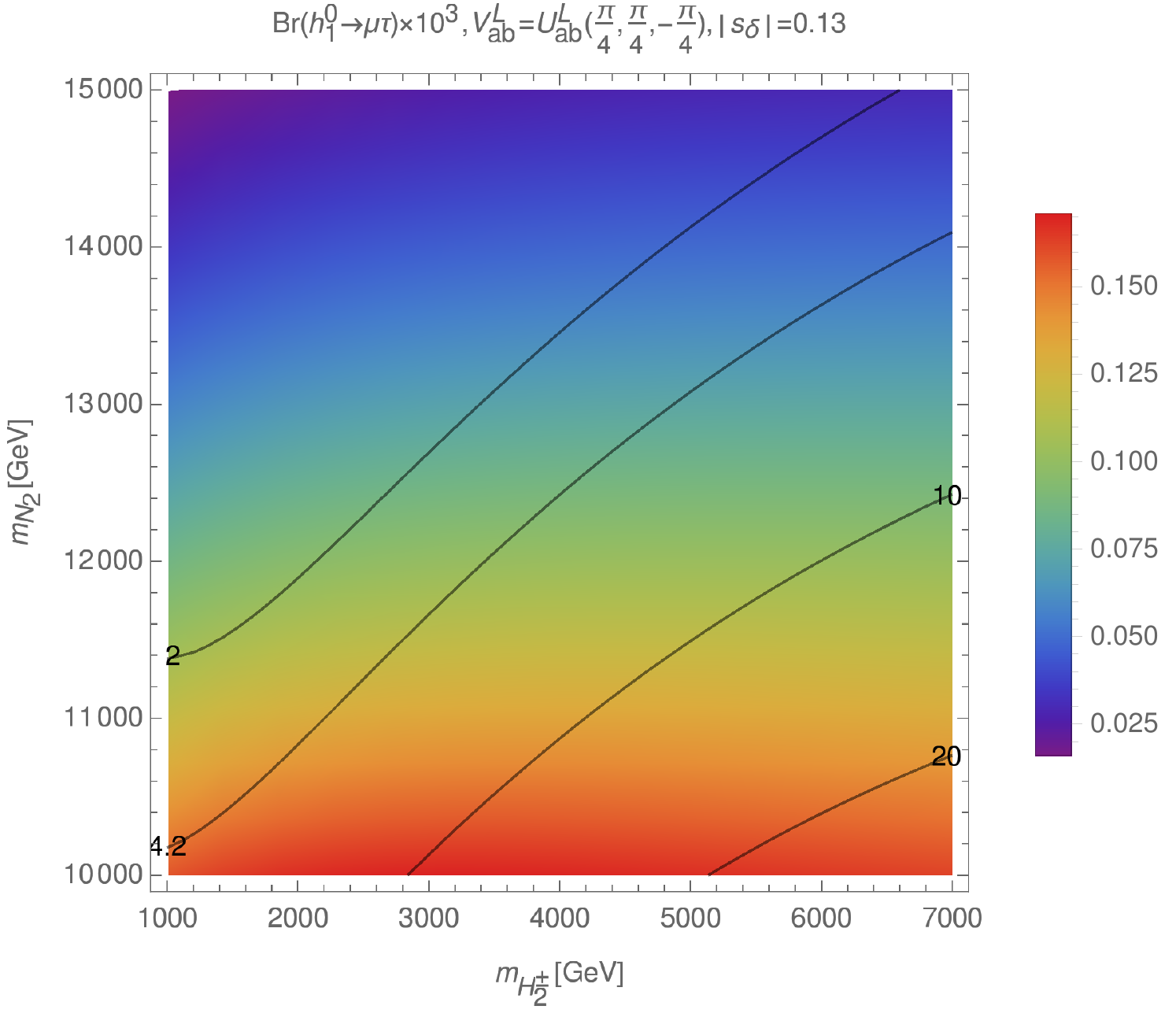}& \includegraphics[width=7.0cm]{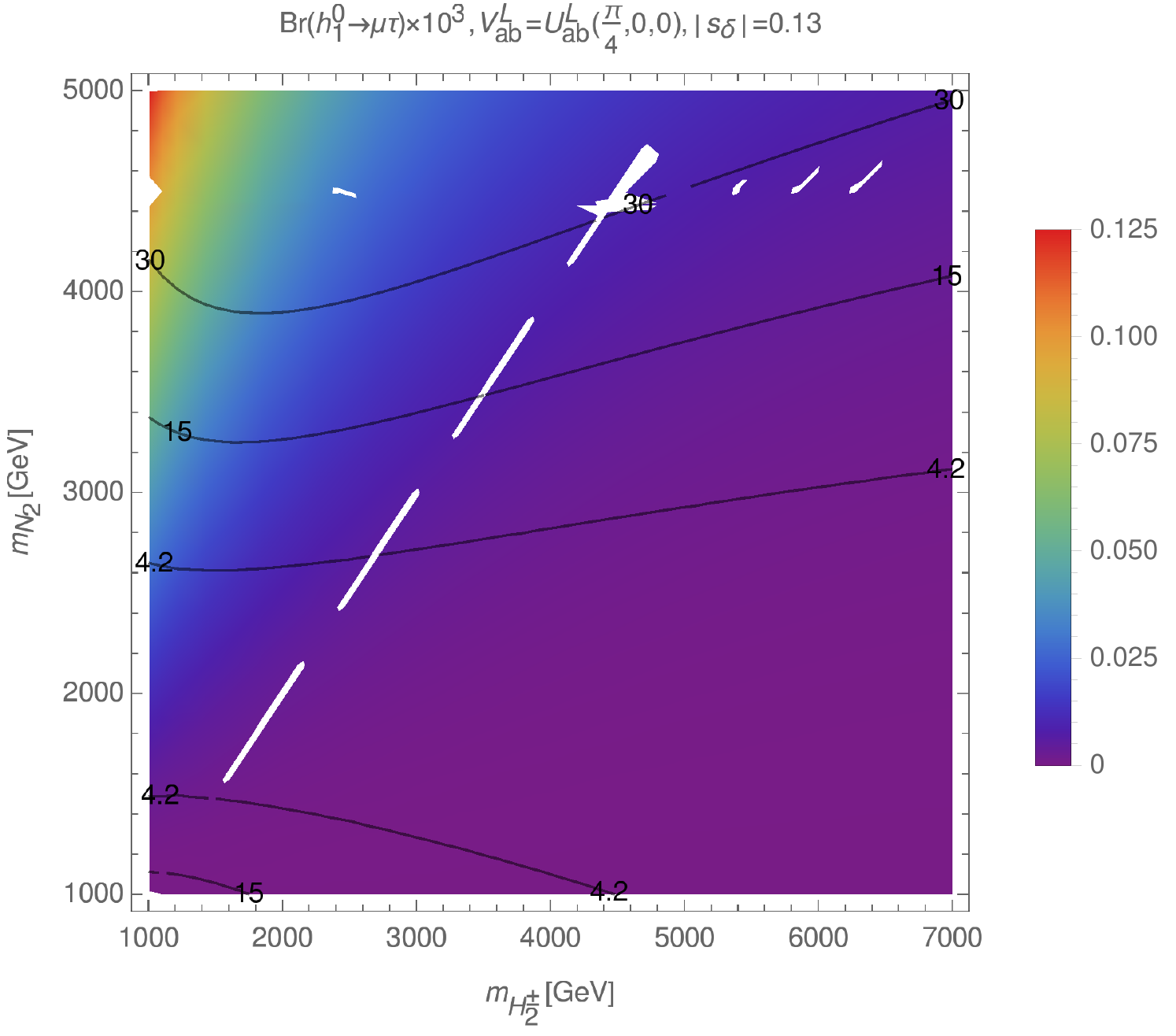}
	\end{tabular}%
	\caption{ Plots of $\mathrm{Br}(h^0_1 \rightarrow \mu\tau)$ depending on $m_{H^\pm_2}$ (first row) and density plots of $\mathrm{Br}(h^0_1 \rightarrow \mu\tau)$ as functions of $m_{H^\pm_2}$ and $m_{N_2}$ (second row) in the case of $V^L_{ab} = U^L_{ab}(\pi/4,\pi/4,-\pi/4)$ (left panel) or in the case of $V^L_{ab} = U^L_{ab}(\pi/4,0,0)$ (right panel).The black cuvers in the second row show the constant values of $\mathrm{Br}(\mu \rightarrow e\gamma)\times 10^{13}$.}
	\label{fig_p23}
\end{figure}

For other cases of mixed matrix exotic leptons, $V^L_{ab} = U^L_{ab}(\pi/4,\pi/4,-\pi/4)$ and $V^L_{ab} = U^L_{ab}(\pi/4,0,0)$, we also show parameter space domains that can give large signals of $\mathrm{Br}(h^0_1 \rightarrow \mu\tau)$ and satisfy the experimental conditions of $(e_i \rightarrow e_j\gamma)$ decays as shown in Fig.(\ref{fig_p23}).

With $V^L_{ab} = U^L_{ab}(\pi/4,0,0)$, $\mathrm{Br}(h^0_1 \rightarrow \mu\tau)$ can achieve to about $10^{-4}$, but in the space domains  satisfying the experimental limits of $(\mu \rightarrow e\gamma)$ decay,  $\mathrm{Br}(h^0_1 \rightarrow \mu\tau)$ can only reach a value less than $10^{-5}$. This result is completely consistent with the corresponding object which was previously published in Ref.\cite{Hue:2015fbb}. When the mixing matrix of exotic leptons has the form $V^L_{ab} = U^L_{ab}(\pi/4,\pi/4,-\pi/4)$, we can obtain allowed parameter space domains that can give signals of $\mathrm{Br}(h^0_1 \rightarrow \mu\tau)$ up to $10^{-4}$. This is the largest signal of $\mathrm{Br}(h^0_1 \rightarrow \mu\tau)$ that we can predict in this model and also very close to the upper limit of this decay ($10^{-3}$) as shown in Refs.\cite{Patrignani:2016xqp,Tanabashi:2018oca,Zyla:2020zbs}. It should be recalled that, we have shown the existence the large signal of $\mathrm{Br}(h^0_1 \rightarrow Z\gamma)$~($1.0\leq \mathrm{R}_{Z\gamma/\gamma \gamma}\leq 2.0$) in Ref.\cite{Hung:2019jue}. Although, the two decays $h^0_1 \rightarrow \mu\tau$ and $h^0_1 \rightarrow Z\gamma$  have different private parts, the common parts are given in the same corresponding form. For example: the common couplings $hV^-V^+ (V^+\equiv W^+,V^+),\, hf\overline{f} (f\equiv e_a),\, hV^-S^+ (S^+\equiv H_{1,2}^+),\, hS^-S^+ (S^+\equiv H_{1,2}^+) $ are given the same form for each decay, the dependent parameters $\la_2, \la_{12}, \la_{13}, \la_{23}$ are given in the same corresponding form and the free parameters such as $\la_1, s_\delta, t_{12}, M_V, ...$ are selected corresponding to the same value domains when examining two decays $h \rightarrow e_ie_j$ and $h \rightarrow Z\gamma$... All common value domains are chosen to be the same. Therefore, we believe that there will exist parameter space domains of this model so that both $h \rightarrow e_ie_j$ and $h \rightarrow Z\gamma$ decays achieve large signals. These are the interested decays of SM-like Higgs boson and their signals are expected to be detectable from large accelerators to confirm the influence of this model.
\section{Conclusions}
\label{conclusion}
The 3-3-1 model with neutral leptons give the Higgs mass spectrum is very diverse when using the Higgs potential in a rather general form as Eq.(\ref{potential}). Applying the same technique as Refs.~\cite{Okada:2016whh,Hung:2019jue}, we can identify two neutral Higgs corresponding to THDM. This leads to model 331NL inheriting some features of THDM as mentioned in Refs.~\cite{Okada:2016whh,Fan:2022dye}. 

We find the contribution of exotic leptons to be the main components for $(e_i \rightarrow e_j\gamma)$ decays. At $13 ~\mathrm{TeV}$ scale of the LHC, leading to  constraints for the masses of some particles such as: $m_V \sim 4.5 ~\mathrm{TeV}$,~ $m_{H_1^\pm} \sim 0.7 ~\mathrm{TeV}$,~ $m_{h_2^0} \sim 1.5 ~\mathrm{TeV}$. By numerical investigation, we show that the parameter space regions satisfying the experimental limits of $(e_i \rightarrow e_j\gamma)$ are highly dependent on the mixing of exotic leptons. However, two cases $V^L_{ab} = U^L_{ab}(\pi/4,\pi/4,\pi/4)$ and $V^L_{ab} = U^L_{ab}(\pi/4,\pi/4,-\pi/4)$ can give roughly the same result when the roles of $m_{N_1}$ and $m_{N_2}$ are swapped. The allowed space regions in this part are all given when fixed at $m_V=4.5 ~\mathrm{TeV}$, exotic leptons have a masses about $2.0 ~\mathrm{TeV}$ or another exotic lepton at $13.0 ~\mathrm{TeV}$.

Although, the forms of the mixing matrix of exotic leptons does not affect the absolute value of the total amplitude of $h^0_1 \rightarrow \mu\tau$ decay , but they affect the regions of parameter space where $\mathrm{Br}(e_i \rightarrow e_j\gamma)$ are satisfied. This suggests us to find the large signal of $\mathrm{Br}(h^0_1 \rightarrow \mu\tau)$  in the allowed space of $e_i \rightarrow e_j\gamma$ decays.

Performing numerical investigation, we show that $\mathrm{Br}(h^0_1 \rightarrow \mu\tau)$ is always less than $10^{-5}$ in the case of $V^L_{ab} = U^L_{ab}(\pi/4,0,0)$ and is in full agreement with previously published results in Ref.~\cite{Hue:2015fbb}. We also show that $\mathrm{Br}(h^0_1 \rightarrow \mu\tau)$ always increases proportionally to $\left| s_\delta \right| $ in all cases of $V^L_{ab}$. Therefore, in the range of values, $0<\left| s_\delta \right|<0.14 $,\, $\mathrm{Br}(h^0_1 \rightarrow \mu\tau)$  can be obtained large values when $\left| s_\delta \right| \rightarrow 0.14 $. Combined with the results shown in Ref.~\cite{Hung:2019jue}, $\mathrm{Br}(h^0_1 \rightarrow Z\gamma)$  can also give large signals in this range of values. So, we can expect to obtain regions of the parameter space for the existence of large signals of both $\mathrm{Br}(h^0_1 \rightarrow \mu\tau)$  and $\mathrm{Br}(h^0_1 \rightarrow Z\gamma)$ in this model. Furthermore, we also predict the large signal of $\mathrm{Br}(h^0_1 \rightarrow \mu\tau)$  can reach $10^{-4}$ in case of $V^L_{ab} = U^L_{ab}(\pi/4,\pi/4,-\pi/4)$. This signal is very close to the upper limit of this channel and is expected to be detectable from large accelerators.
\section*{Acknowledgments}
This research is funded by Vietnam National Foundation for Science and Technology Development (NAFOSTED) under grant number 103.01-2020.01.

\appendix
\section{Higgs and gauge bosons in the 331NL model.}
\label{appen_HGB}

{\bf Higgs bosons}

From Eq.~(\ref{potential}), we have the minimum conditions of the Higgs potential as, 
\begin{eqnarray}
	&&\mu _1^2 = \frac{f v_1 v_3^2}{ v_2}-\frac{\lambda _{12} v_1^2 +\lambda _{13} v_3^2 }{2}-\lambda _1 v_2^2,\nonumber \\
	&&\mu _2^2 = \frac{f v_2 v_3^2}{ v_1}-\frac{\lambda _{12} v_2^2 +\lambda _{23} v_3^2 }{2}-\lambda _2 v_1^2,
	\nonumber \\
	&&
	\mu _3^2 = f v_2 v_1-\lambda _3 v_3^2-\frac{ \left(\lambda _{23}
		v_1^2+\lambda _{13} v_2^2\right)}{2}.
	\label{mincond} 
\end{eqnarray}

There are two Goldstone bosons $G^{\pm}_W$ and $G^{\pm}_V$ of the respective singly charged gauge bosons $W^{\pm}$ and $V^{\pm}$. Two other massive singly charged Higgses have masses
	\bea 	m_{{H_1}^{\pm }}^2 = \left(v_1^2+v_2^2\right) \left(\frac{\tilde{\lambda }_{12}}{2}+\frac{f
		v_{3}^2}{v_1v_2}\right);\hs m_{{H_2}^{\pm }}^2 = \left(v_1^2+v_3^2\right) \left(\frac{\tilde{\lambda
		}_{23}}{2}+\frac{fv_2}{v_{1}}\right). \label{massedcH}
	\eea

The relation between two  flavor and mass bases of the singly charged Higgses are
	\bea \left(
		\begin{array}{c}
			\rho^{\pm} \\
			\eta^{\pm} \\
		\end{array}
		\right)= \left(
		\begin{array}{cc}
			-c_{12} & s_{12} \\
			s_{12}& c_{12} \\
		\end{array}
		\right) \left(
		\begin{array}{c}
			G^{\pm}_W \\
			H_1^{\pm} \\
		\end{array}
		\right), \hs  \left(
		\begin{array}{c}
			\rho'^{\pm} \\
			\chi^{\pm} \\
		\end{array}
		\right)=\left(
		\begin{array}{cc}
			-s_{13} & c_{13} \\
			c_{13}& s_{13} \\
		\end{array}
		\right) \left(
		\begin{array}{c}
			G^{\pm}_V \\
			H_2^{\pm} \\
		\end{array}
		\right),
		\label{sHiggse}\eea
		where $s_{ij}\equiv\sin\beta_{ij}$,~ $c_{ij}\equiv\cos\beta_{ij}$, and $t_{12}\equiv\tan\beta_{12}=\frac{v_2}{v_1},\,t_{13}\equiv\tan\beta_{13}=\frac{v_1}{v_3},\,t_{23}\equiv\tan\beta_{23}=\frac{v_2}{v_3}$.
	
 With the components of selected scalar fields as Eq.(\ref{vevs}), we obtain $5$ real scalars initially, namely $S_1, S_2, S_3, S_2'. S_3'$. In the final state, similar as Refs.~\cite{Hue:2015mna,Hue:2015fbb}, we get 4 massive Higgses and a Goldstone boson ($G_{U}$) corresponding to the gauge boson $U$. A heavy neutral Higgs mixed with $G_{U}$ on the original basis ($S'_2, S'_3$) is:
	
	\bea   \left(
	\begin{array}{c}
		S'_{2} \\
		S'_{3} \\
	\end{array}
	\right)=\left(
	\begin{array}{cc}
		-s_{13} & c_{13} \\
		c_{13}& s_{13} \\
	\end{array}
	\right) \left(
	\begin{array}{c}
		G_U \\
		h_4^{0} \\
	\end{array}
	\right),
	\label{nrHigg1}\eea
	and mass of $h^0_4$ was given:
	\bea
	m^2_{h_4^0}= \left(v_1^2+v_3^2\right) \left(\frac{\tilde{\lambda }_{13}}{2}+\frac{fv_2}{ v_{1}}\right)
	\eea
	The remainders are three neutral Higgses whose mass mixing matrix on the flavor basis ($S_1,S_2,S_3$) is:
	\bea
	\mathcal{M}^2_h=\left(
	\begin{array}{ccc}
		2 \lambda _2 v_1^2+\frac{f v_2 v_3^2}{v_1} & v_1 v_2 \lambda _{12}-f v_3^2 & v_3 \left(v_1 \lambda _{23}-f v_2\right)  \\
		v_1 v_2 \lambda _{12}-f v_3^2 & 2 \lambda _1 v_2^2+\frac{f v_1 v_3^2}{v_2} & v_3 \left(v_2 \lambda _{13}-f v_1\right) \\
		v_3 \left(v_1 \lambda _{23}-f v_2\right) & v_3 \left(v_2 \lambda _{13}-f v_1\right) & 2 \lambda _3 v_3^2+f v_1 v_2  \\
	\end{array}
	\right)\label{massenH}
	\eea
	Among of the three neutral Higgses mentioned in Eq.(\ref{massenH}), the lightest $h^0_1$ is identified with the Higgs boson in the standard model, called: SM-like Higgs boson. To avoid the tree level contributions of SM-like Higgs boson to the  flavor  changing neutral currents (FCNC) in the quark sector, we used the aligned limit introduced in Refs.~\cite{Okada:2016whh,Hung:2019jue}, namely
	\begin{equation}\label{eq_alignH0}
			f=\lambda_{13}t_{12} =\frac{\lambda_{23}}{t_{12}}.
	\end{equation}	
	For simplicity, we choose $f$ and $\lambda_{23}$ as functions of the remaining. Thus, the mass matrix of the Higgses at Eq.(\ref{massenH}) now becomes
	\bea
		\left(
		\begin{array}{ccc}
			2 \lambda _2 v_1^2+\lambda _{13}v_3^2t_{12}^2 & (\lambda _{12}v_1^2- \lambda	_{13}v_3^2)t_{12}  & 0 \\
			(\lambda _{12}v_1^2- \lambda_{13}v_3^2)t_{12} & 2\lambda _1 v_2^2+\lambda _{13}v_3^2 & 0 \\
			0 & 0 & 2 \lambda _3 v_3^2+\lambda _{13}v_2^2\\
		\end{array}
		\right)\label{eq_m2r}
		\eea
	
	As a result, $S_3\equiv h^0_3$ is a physical CP-even neutral Higgs boson with mass $m^2_{h^0_3}= \lambda_{13}v_2^2 +2\lambda_3 v_3^2 $.  The sub-matrix $2\times 2$ in Eq.~\eqref{eq_m2r} is denoted as $M'^2_h$, which is diagonalized as follows,
	\begin{align}\label{eq-mixingh0}
		R(\alpha)M'^2_hR^T(\alpha)= \mathrm{diag}(m^2_{h^0_1},  m^2_{h^0_2}),
	\end{align}
	where  
	\begin{align}
		\alpha&\equiv \beta_{12}  -\frac{\pi}{2} + \delta~ \mathrm{and} ~R(\alpha) =\left(
		\begin{array}{cc}
			c_{\al} & -s_{\al} \\
			s_{\al}& c_{\al} \\
		\end{array}
		\right), \label{eq_alpha}
	\end{align}
	Using the techniques described in Refs.~\cite{Okada:2016whh,Hung:2019jue}, we obtain that the masses of neutral Higgses depend on the mixing angle $\delta$ - this is a characteristic parameter for THDM. As mentioned in Ref.~\cite{Kanemura:2018yai}, this parameter constraints $c_\delta >0.99$ for all THDMs, resulting in $\left| s_\delta \right| <0.14$.
	\begin{align}
		m^2_{h^0_1}&= M^2_{22}\cos^2\delta +M^2_{11}\sin^2\delta - M^2_{12}\sin2\delta,\crn
		m^2_{h^0_2}&=  M^2_{22}\sin^2\delta +M^2_{11}\cos^2\delta + M^2_{12}\sin2\delta, \crn
		\tan2\delta&= \frac{2M^2_{12}}{M^2_{22} -M^2_{11}}. \label{ang_mixingh0}
	\end{align}
	The components $M_{\mathrm{ij}}$ of a $2\times 2$ matrix are formed from the sub-matrix of $\mathcal{M}^2_h$ after rotating the angle $\beta_{12}$.
	\begin{align}
		M^2_{11}&= 2s^2_{12}c^2_{12}\left[ \lambda_1 +\lambda_2 -\lambda_{12}\right] v^2 +\frac{\lambda_{13}v^2_3}{c^2_{12}},\crn
		M^2_{12}&= \left[\lambda_1 s^2_{12} - \lambda_2 c^2_{12} -\lambda_{12} (s^2_{12} -c^2_{12})\right]s_{12}c_{12} v^2 =\mathcal{O}(v^2),\crn
		M^2_{22}&= 2\left( s_{12}^4 \lambda_{1} +c_{12}^4 \lambda_{2} +s_{12}^2 c_{12}^2 \lambda_{12}\right) v^2=\mathcal{O}(v^2),~v^2=v_1^2+v_2^2.  \label{eq_mixingh0} 
	\end{align}
	We also have 
	\begin{align}
		\left(
		\begin{array}{c}
			S_2 \\
			S_1\\
		\end{array}
		\right)&= R^T(\alpha)\left(
		\begin{array}{c}
			h^0_1 \\
			h^0_2 \\
		\end{array}
		\right). \label{mixingh0}
	\end{align}
	The lightest $h^0_1$ is SM-like Higgs boson found at LHC. From Eqs.(\ref{ang_mixingh0},\ref{eq_mixingh0}), we can see that  $\tan2\delta= \frac{2M^2_{12}}{M^2_{22} -M^2_{11}}=\mathcal{O}(\frac{v^2}{v_3^2}) \simeq0$ when $v^2\ll v_3^2$. In this limit, $m^2_{h}= M^2_{22} + v^2\times \mathcal{O}(\frac{v^2}{v_3^2})\sim  M^2_{22}$ while $m^2_{h^0_2}= M^2_{11} + v^2 \times  \mathcal{O}(\frac{v^2}{v_3^2}) \simeq M^2_{11}$. In the next section, we will see more explicitly that the couplings of $h_1^0$ are the same as those given in the SM in the limit $\delta\rightarrow 0$. 
	
	Using the invariance trace of the squared mass matrices in Eq.(\ref{eq-mixingh0}), we have
       \bea
		2 \lambda _2 v_1^2+\lambda _{13}v_3^2t_{12}^2+2\lambda _1 v_2^2+\lambda _{13}v_3^2=m^2_{h^0_1}+m^2_{h^0_2}
		\eea
	the $\lambda_{13}$ can be written as 
	\begin{equation}
			\lambda_{13}=\frac{c^2_{12}}{v_3^2} \left[  m^2_{h^0_1} +m^2_{h^0_2} -\frac{8m_W^2}{g^2}\left( \lambda_1s_{12}^2 +\lambda_2c_{12}^2  \right)\right] . 
	\end{equation}
	The other Higgs self couplings have given in Tab.\ref{albga}. They should satisfy all constraints discussed in the literature to guarantee the pertubative limits, the vacuum stability of the Higgs potential~\cite{Sanchez-Vega:2018qje}, and the positive squared masses of all Higgs bosons.
	
{\bf Gauge bosons}

The $SU(3)_L\otimes U(1)_X$ includes 8 generators $T^a$ (a=1,8) of the $SU(3)_L$ and a generator $T^9$  of the $U(1)_X$, corresponding to eight gauge bosons $W^a_{\mu}$  and the $X_{\mu}$ of the $U(1)_X$ . The respective covariant derivative is
\be D_{\mu}\equiv \partial_{\mu}-i g_3 W^a_{\mu}T^a-g_1 T^9X X_{\mu}. \label{code} \ee
The Gell-Mann matrices are denoted as $\lambda_a$, we have $T^a=\frac{1}{2}\lambda_a,-\frac{1}{2}\lambda_a^T$  or $0$ depending on the triplet, antitriplet or singlet representation of the $SU(3)_L$ that  $T^a$ acts on. The $T^9$ is defined as $T^9=\frac{1}{\sqrt{6}}$ and $X$ is the $U(1)_X$ charge of the field it acts on.
We also have defined $W_\mu^+ = \frac{1}{\sqrt{2}}(W_\mu^1-iW_\mu^2)$, as usual, $V_\mu^-=\frac{1}{\sqrt{2}}(W_\mu^6-iW_\mu^7)$ and $U_\mu^0=\frac{1}{\sqrt{2}}(W_\mu^4-iW_\mu^5)$: 

\bea W^a_{\mu} T^a= \frac{1}{\sqrt{2}} \left(
\begin{array}{ccc}
	0 & W^{+}_{\mu} & U^0_{\mu} \\
	W^{-}_{\mu} & 0 & V^{-}_{\mu} \\
	U^{0*}_{\mu} & V^{+}_{\mu} & 0 \\
\end{array}
\right).
\label{gaugeboson1} \eea
The masses of these gauge bosons are:
\be  m_W^2=\frac{g^2}{4}\left(v^2_1+ v^2_2\right),\hs m^2_{U}=\frac{g^2}{4}\left(v^2_2+ v^2_3\right),\hs m^2_{V} =\frac{g^2}{4}\left(v^2_1+v^2_3\right), \label{gmass}\ee
where we used the relation  $v_1^2+v_2^2=v^2\equiv 246^2 \mathrm{GeV^2}$ so that the mass of the W-boson in the 331NL model matches the corresponding one in the SM. 

The three remaining neutral gauge bosons, $A_\mu$, $Z_\mu$ and $Z^\prime_\mu$, couple to the fermions in a diagonal basis as shown in Ref.~\cite{Hung:2019jue}. These couplings do not correlate with LFV decays, so we do not mention them in this work.
\section{Master integrals.}
\label{appen_PV}
To calculate the contributions at one-loop order of the Feynman diagrams in Figures \ref{fig_lalbga} and \ref{fig_hmt331}, we use the Passarino-Veltman (PV) functions as mentioned in Ref.\cite{Passarino:1978jh}. By introducing the notations $D_0=k^2-M_0^2+i\delta$, $D_1=(k-p_1)^2-M_{1}^2+i\delta$ and $D_2=(k+p_2)^2-M_2^2+i\delta$, where $\delta$ is  infinitesimally a  positive real quantity, we have:
\bea
A_{0}(M_n)
&\equiv &\frac{\left(2\pi\mu\right)^{4-D}}{i\pi^2}\int \frac{d^D k}{D_n}, \hs
B^{(1)}_0 \equiv\frac{\left(2\pi\mu\right)^{4-D}}{i\pi^2}\int \frac{d^D k}{D_0D_1},\crn
B^{(2)}_0 &\equiv &\frac{\left(2\pi\mu\right)^{4-D}}{i\pi^2}\int \frac{d^D k}{D_0D_2}, \hs
B^{(12)}_0 \equiv \frac{\left(2\pi\mu\right)^{4-D}}{i\pi^2}\int \frac{d^D k}{D_1D_2},\crn
C_0&\equiv&  C_{0}(M_0,M_1,M_2) =\frac{1}{i\pi^2}\int \frac{d^4 k}{D_0D_1D_2},
\label{scalrInte}\eea
where $n=1,2$, $D=4-2\epsilon \leq 4$ is the dimension of the integral, while $~M_0,~M_1,~M_2$ stand for the masses of virtual particles in the loops. We also assume  $p^2_1=m^2_{1},~p^2_2=m^2_{2}$ for external fermions. The tensor integrals are
\bea
A^{\mu}(p_n;M_n)
&=&\frac{\left(2\pi\mu\right)^{4-D}}{i\pi^2}\int \frac{d^D k\times k^{\mu}}{D_n}=A_0(M_n)p_n^{\mu},\crn
B^{\mu}(p_n;M_0,M_n)&=& \frac{\left(2\pi\mu\right)^{4-D}}{i\pi^2}\int \frac{d^D k\times
	k^{\mu}}{D_0D_n}\equiv B^{(n)}_1p^{\mu}_n,\crn
B^{\mu}(p_1,p_2;M_1,M_2)&=& \frac{\left(2\pi\mu\right)^{4-D}}{i\pi^2}\int \frac{d^D k\times
	k^{\mu}}{D_1D_2}\equiv B^{(12)}_1p^{\mu}_1+B^{(12)}_2p^{\mu}_2,\crn
C^{\mu}(M_0,M_1,M_2)&=&\frac{1}{i\pi^2}\int \frac{d^4 k\times k^{\mu}}{D_0D_1D_2}\equiv  C_1 p_1^{\mu}+C_2 p_2^{\mu},\crn
C^{\mu \nu}(M_0,M_1,M_2)&=&\frac{1}{i\pi^2}\int \frac{d^4 k\times k^{\mu}k^{\nu}}{D_0D_1D_2}\equiv  C_{00}g^{\mu \nu}+C_{11} p_1^{\mu}p_1^{\nu}+C_{12} p_1^{\mu}p_2^{\nu}+C_{21} p_2^{\mu}p_1^{\nu}+C_{22} p_2^{\mu}p_2^{\nu},\crn
\label{oneloopin1}\eea
where $A_0$, $B^{(n)}_{0,1}$, $B^{(12)}_{n}$ and $C_{0,n}, C_{mn}$   are PV functions.  It is well-known that $C_{0,n}, C_{mn}$ are finite while the remains are divergent. We denote
\be \Delta_{\epsilon}\equiv \frac{1}{\epsilon}+\ln4\pi-\gamma_E, \label{divt}\ee with $\gamma_E$ is the  Euler constant.  

Using the technique as mentioned in Ref.\cite{Hue:2017lak}, we can show the divergent parts of the above PV functions as
\bea  \mathrm{Div}[A_0(M_n)]&=& M_n^2 \Delta_{\epsilon}, \hs  \mathrm{Div}[B^{(n)}_0]= \mathrm{Div}[B^{(12)}_0]= \Delta_{\epsilon}, \crn
\mathrm{Div}[B^{(1)}_1]&=&\mathrm{Div}[B^{(12)}_1] = \frac{1}{2}\Delta_{\epsilon},  \hs  \mathrm{Div}[B^{(2)}_1] = \mathrm{Div}[B^{(12)}_2]= -\frac{1}{2} \Delta_{\epsilon}.  \label{divs1}\eea
Apart from the divergent parts, the rest of these functions are finite.

Thus, the above PV functions can be written in form:
\be  A_0(M)= M^2\Delta_{\epsilon}+a_0(M),\,\, B^{(n)}_{0,1}= \mathrm{Div}[B^{(n)}_{0,1}]+ b^{(n)}_{0,1}, \,\,  B^{(12)}_{0,1,2}= \mathrm{Div}[B^{(12)}_{0,1,2}]+ b^{(12)}_{0,1,2}, \label{B01i}\ee
where $a_0(M), \,\,b^{(n)}_{0,1}, \,\, b^{(12)}_{0,1,2} $ are finite parts and have a specific form defined as Ref.\cite{Hue:2017lak} for $e_i \rightarrow e_j\ga$ decays and Ref.\cite{Thuc:2016qva} for $h^0_1 \rightarrow \mu \tau$ decay.

\section{Analytic formulas of one-loop order  for $e_i \rightarrow e_j\ga$ decays.}
\label{appen_loops1}
We use techniques as shown in \cite{Hue:2017lak,Hung:2021fzb} to give the factors at one-loop order of the $e_i \rightarrow e_j\gamma$ decays. The Passarino-Veltman functions obey the rules as shown in \cite{Passarino:1978jh}, and have a common set of variables ($p_k^2,m_1^2,m_2^2,m_3^2$) with $p_k^2=m_{e_i}^2,0,m_{e_j}^2$ related to external momenta and $m_1^2,m_2^2,m_3^2$ related to masses in loop of figures \ref{fig_lalbga}. For brevity, we use the notations: $C_{0,n} \equiv C_{0,n}(p_k^2,m_1^2,m_2^2,m_3^2)$ and $C_{mn} \equiv C_{mn}(p_k^2,m_1^2,m_2^2,m_3^2); m,n=1,2$ in the analytic formulas below. 

Factors of diagram (1) of Figure \ref{fig_lalbga}.
\bea
\mathcal{D}^{\nu_a WW}_{(ij)L}(m_{\nu_a}^2,m_W^2) &= -\frac{eg^2m_{e_j}}{32\pi^2} \left[2(C_1 + C_{12} + C_{22}) + \fr{m_{e_i}^2}{m_W^2}(C_{11} + C_{12} - C_1) \right.\crn
& \left. + \fr{m_{\nu_a}^2}{m_W^2}(C_0 + C_{12} + C_{22} - C_1 - 2C_2)\right],\label{eq:DL_nuWW}\\
\mathcal{D}_{(ij)R}^{\nu_a WW}(m_{\nu_a}^2,m_W^2) &= -\frac{eg^2m_{e_i}}{32\pi^2} \left[2(C_2 + C_{11} + C_{12}) + \fr{m_{e_j}^2}{m_W^2}(C_{12} + C_{22} - C_2) \right.\crn
& \left. + \fr{m_{\nu_a}^2}{m_W^2}(C_0 + C_{11} + C_{12} - 2C_1 - C_2)\right],
\label{eq:DR_nuWW}
\eea

Factors of diagram (2) of Figure \ref{fig_lalbga}.
\bea
\mathcal{D}^{N_aVV}_{(ij)L}(m_{Na}^2,m_V^2) &=-\frac{eg^2m_{e_j}}{32\pi^2} \left[2(C_1 + C_{12} + C_{22}) + \fr{m_{e_i}^2}{m_V^2}(C_{11} + C_{12} - C_1) \right.\crn
& \left. + \fr{m_{N_a}^2}{m_V^2}(C_0 + C_{12} + C_{22} - C_1 - 2C_2)\right],\label{eq:DL_NaVV}\\
\mathcal{D}_{(ij)R}^{N_aVV}(m_{Na}^2,m_V^2) &= -\frac{eg^2m_{e_i}}{32\pi^2} \left[2(C_2 + C_{11} + C_{12}) + \fr{m_{e_j}^2}{m_V^2}(C_{12} + C_{22} - C_2) \right.\crn
& \left. + \fr{m_{N_a}^2}{m_V^2}(C_0 + C_{11} + C_{12} - 2C_1 - C_2)\right],
\label{eq:DR_NaVV}
\eea
Factors of diagram (3) of Figure \ref{fig_lalbga}.
\bea
\mathcal{D}_{(ij)L}^{\nu_a H_1H_1}(m_{\nu_a}^2,m_{H_1}^2) &=-\frac{eg^2m_{e_j}}{64\pi^2}\left[\fr{m_{e_i}^2}{m_W^2}(C_{11} + C_{12} - C_1) + \fr{m_{\nu_a}^2}{m_W^2}(C_{12} + C_{22} - C_2) \right.\crn
& \left. + \fr{m_{\nu_a}^2}{m_W^2}(C_1 + C_2 - C_0)\right],\label{eq:DL_nuH_1H_1}\\
\mathcal{D}_{(ij)R}^{\nu_a H_1 H_1}(m_{\nu_a}^2,m_{H_1}^2) &=-\frac{eg^2m_{e_i}}{64\pi^2}\left[\fr{m_{e_j}^2}{m_W^2}(C_{12} + C_{22} - C_2) + \fr{m_{\nu_a}^2}{m_W^2}(C_{11} + C_{12} - C_1) \right.\crn
& \left. + \fr{m_{\nu_a}^2}{m_W^2}(C_1 + C_2 - C_0)\right],
\label{eq:DR_nuH1H1}
\eea
Factors of diagram (4) of Figure \ref{fig_lalbga}.
\bea
\mathcal{D}_{(ij)L}^{Na H_2H_2}(m_{N_a}^2,m_{H_2}^2) &=-\frac{eg^2m_{e_j}}{32\pi^2}\left[\fr{m_{e_i}^2}{m_V^2}(C_{11} + C_{12} - C_1) + \fr{m_{N_a}^2}{m_V^2}(C_{12} + C_{22} - C_2) \right.\crn
& \left. + \fr{m_{N_a}^2}{m_V^2}(C_1 + C_2 - C_0)\right],\label{eq:DL_NaH_2H_2}\\
\mathcal{D}_{(ij)R}^{Na H_2H_2}(m_{N_a}^2,m_{H_2}^2) &=-\frac{eg^2m_{e_i}}{32\pi^2}\left[\fr{m_{e_j}^2}{m_W^2}(C_{12} + C_{22} - C_2) + \fr{m_{N_a}^2}{m_V^2}(C_{11} + C_{12} - C_1) \right.\crn
& \left. + \fr{m_{N_a}^2}{m_V^2}(C_1 + C_2 - C_0)\right],
\label{eq:DR_NaH2H2}
\eea

\section{Analytic formulas of one-loop order  for $h^0_1 \rightarrow e_ie_j$ decays.}
\label{appen_loops2}
The one-loop factors of the diagrams in Fig.(\ref{fig_hmt331}) are given in this appendix. We used the same calculation techniques as shown in \cite{Thuc:2016qva,Phan:2016ouz}. We denote $m_{e_i} \equiv  m_1$ and $ m_{e_j} \equiv m_2$.
{\small\bea  \mathcal{M}^{FVV}_L(m_F,m_V)
	&=&m_V m_1\left\{ \frac{1}{2m_V^4}\left[m_F^2(B^{(1)}_1-B^{(1)}_0-B^{(2)}_0)\right.\right.\crn
	\hs &-&\left.\left. m_2^2B^{(2)}_1 + \left(2m_V^2+m^2_{h^0}\right)m_F^2\left(C_0-C_1\right)\right]\right.\crn
	&&\left.-\left(2+\frac{m_1^2-m_2^2}{m_V^2}\right) C_1 +
	\left(\frac{m_1^2-m^2_{h^0}}{m_V^2}+ \frac{m_2^2 m^2_{h^0}}{2m_V^4}\right)C_2\right\}, \label{EfvvL} \\
	\mathcal{M}^{FVV}_R(m_F,m_V)&=&m_V m_2\left\{\frac{1}{2 m_V^4}\left[-m_F^2\left(B^{(2)}_1+ B^{(1)}_0 + B^{(2)}_0 \right) \right.\right.\crn
	&+& \left.\left.  m_1 ^2 B^{(1)}_1  +   (2m_V^2+m^2_{h^0}) m_F^2(C_0+C_2)\right] \right.\crn
	&&\left.+\left(2+\frac{-m_1^2+m_2^2}{m_V^2}\right)C_2-\left( \frac{m_2^2-m^2_{h^0}}{m_V^2}+ \frac{m_1^2 m^2_{h^0}}{m_V^4}\right)C_1\right\},  \label{EfvvR}
	\eea
	\bea
	&& \mathcal{M}^{FVH}_L(a_1,a_2,v_1,v_2,m_F,m_V,m_H)\crn&=&
	m_1\left\{-\fr{a_2}{v_2} \fr{m_F^2}{m_V^2}\left(B^{(1)}_1-B^{(1)}_0\right)   + \fr{a_1}{v_1}m_2^2\left[2 C_1-\left(1+ \fr{m^2_{h}-m^2_{h^0}}{m_V^2}\right) C_2\right]\right.\crn
	&&\left.+\fr{a_2}{v_2}m_F^2\left[C_0+C_1+\fr{m^2_{h}-m^2_{h^0}}{m_V^2}\left(C_0-C_1\right) \right]\right\}, \label{EfvhL} \\
	&& \mathcal{M}^{FVH}_Ra_1,a_2,v_1,v_2,m_F,m_V,m_H)\crn &=&m_2\left\{\fr{a_1}{v_1}\left[\fr{m_1^2B^{(1)}_1-m_F^2B^{(1)}_0}{m_V^2} +\left(\frac{}{}m_F^2C_0-m_1^2C_1+2 m_2^2C_2\right.\right.\right.\crn
	&&\left.\left.+2(m^2_{h^0}-m_2^2)C_1-  \fr{m^2_{h}-m^2_{h^0}}{m_V^2}\left(m^2_FC_0-m_1^2C_1\right)\right)\right]\crn &&+\left.\fr{a_2}{v_2} m_F^2\left(-2C_0-C_2+\fr{m^2_{h}-m^2_{h^0}}{m_V^2}C_2 \right) \right\},   \label{EfvhR}
	\eea
	\bea
	&& \mathcal{M}^{FHV}_L(a_1,a_2,v_1,v_2,m_F,m_H,m_V)\crn&=& m_1\left\{\fr{a_1}{v_1}\left[\fr{-m_2^2B^{(2)}_1-m_F^2B^{(2)}_0}{m_V^2} +\left(\frac{}{}m_F^2C_0-2m_1^2C_1+ m_2^2C_2\right.\right.\right.\crn
	&&\left.\left.-2(m^2_{h^0}-m_1^2)C_2-  \fr{m^2_{h}-m^2_{h^0}}{m_V^2}\left(m^2_FC_0+m_2^2C_2\right)\right)\right] \crn &&\left.+\fr{a_2}{v_2} m_F^2\left(-2C_0+C_1-\fr{m^2_{h}-m^2_{h^0}}{m_V^2}C_1 \right)\right\}, \label{EfvhL} \\
	&& \mathcal{M}^{FHV}_R(a_1,a_2,v_1,v_2,m_F,m_H,m_V)\crn&=& m_2 \left\{\fr{a_2}{v_2} \fr{m_F^2}{m_V^2}\left(B^{(2)}_1+B^{(2)}_0\right)
	+ \fr{a_1}{v_1}m_1^2\left[-2 C_2+\left(1+ \fr{m^2_{h}-m^2_{h^0}}{m_V^2}\right) C_1\right]\right.\crn
	&&\left.+\fr{a_2}{v_2}m_F^2\left[C_0-C_2+\fr{m^2_{h}-m^2_{h^0}}{m_V^2}\left(C_0+C_2\right) \right]\right\}.   \label{EfvhR}
	\eea
	\bea
	\mathcal{M}^{FV}_L(m_F,m_V)&=& \fr{-m_1m_2^2}{m_V(m_1^2-m_2^2)}\left[\left(2+\frac{m_F^2}{m_V^2}\right) \left(B^{(1)}_1 +B^{(2)}_1 \right) \right. \crn&+&\left.\fr{m_1^2 B^{(1)}_1 +m_2^2 B^{(2)}_1}{m_V^2} - \fr{2m_F^2}{m_V^2}\left(B^{(1)}_0-B^{(2)}_0\right)\right],  \label{DfvL} \\
	\mathcal{M}^{FV}_R(m_F,m_V)&=& \frac{m_1}{m_2}E^{FV}_L, \label{DfvR}\eea
	\bea
	\mathcal{M}^{HFF}_L(a_1,a_2,v_1,v_2,m_F,m_H)&=&\frac{ m_1m^2_F }{v_2}\crn
	&\times& \left[\dfrac{a_1a_2}{v_1v_2}B^{(12)}_{0}
	+\fr{a_1^2}{v_1^2}m_2^2(2C_2+C_0)+\fr{a_2^2}{v_2^2}m_F^2(C_0-2C_1) \right.\crn
	&+& \left.\fr{a_1a_2}{v_1v_2} \left(\frac{}{}2m_2^2C_2-(m_1^2+m_2^2)C_1+(m_F^2+m^2_{h}+m_2^2)C_0\right)\right],\crn  \label{EhffL} \\
	\mathcal{M}^{HFF}_R(a_1,a_2,v_1,v_2,m_F,m_H)&=& \frac{m_2 m^2_F}{v_2}\crn
	&\times&\left[ \dfrac{a_1a_2}{v_1v_2}B^{(12)}_{0}+ \dfrac{a_1^2}{v_1^2}m_1^2(C_0-2C_1)+\fr{a_2^2}{v_2^2}m_F^2(C_0+2C_2)\right.\crn
	&+&\left. \fr{a_1a_2}{v_1v_2}\left(\frac{}{}-2m_1^2C_1+(m_1^2+m_2^2)C_2+(m_F^2+m^2_{h}+m_1^2)C_0 \right)\right], \crn\label{EhffR} \eea
	\bea
	\mathcal{M}^{FHH}_L(a_1,a_2,v_1,v_2,m_F,m_H)&=&  m_1v_2\left[ \fr{a_1a_2}{v_1v_2}m_F^2C_0-\fr{a^2_1}{v^2_1}m_2^2C_2+\fr{a^2_2}{v^2_2}m_F^2C_1\right] , \crn \label{EfhhL} \\
	\mathcal{M}^{FHH}_R(a_1,a_2,v_1,v_2,m_F,m_H)&=& m_2v_2\left[ \fr{a_1a_2}{v_1v_2}m_F^2C_0+\fr{a^2_1}{v^2_1}m_1^2C_1-\fr{a^2_2}{v^2_2}m_F^2C_2 \right],\crn \label{EfhhR} \eea
	\bea
	\mathcal{M}^{VFF}_L(m_V,m_F)&=&\frac{m_1m^2_F}{m_V}\crn
	&\times&\left[\fr{1}{m_V^2}\left(B^{(12)}_{0}
	+B^{(1)}_1 -(m_1^2+m_2^2-2m_F^2)C_1\right)-C_0+4C_1\right],\crn  \label{EvffL} \\
	\mathcal{M}^{VFF}_R(m_V,m_F)&=&
	\frac{m_2 m^2_F}{m_V} \crn
	&\times&\left[ \fr{1}{m_V^2}\left(B^{(12)}_{0} -B^{(2)}_1 +(m_1^2+m_2^2-2m_F^2)C_2\right)-C_0-4C_2 \right],\crn \label{EvffR}\eea

	\bea
	\mathcal{M}^{FH}_L(a_1,a_2,v_1,v_2,m_F,m_H)&=& \fr{m_1}{v_1(m_1^2-m_2^2)}\left[m^2_2 \left(m^2_1\fr{a_1^2}{v_1^2}+m^2_F\fr{a^2_2}{v^2_2} \right)
	\left(B_1^{(1)}+B_1^{(2)}\right) \right.\crn
	&&\left.  \hspace{1.8 cm}+m^2_F\fr{a_1a_2}{v_1v_2}\left(2m^2_2B_0^{(1)}-(m^2_1 +m^2_2)B_0^{(2)}\right) \right], \label{DfhL} \\
	\mathcal{M}^{FH}_R(a_1,a_2,v_1,v_2,m_F,m_H)&=&  \fr{m_2}{v_1(m_1^2-m_2^2)}\left[ m^2_1 \left(m^2_2\fr{a_1^2}{v_1^2}+m^2_F\fr{a^2_2}{v^2_2} \right)\left(B_1^{(1)}+B_1^{(2)}\right)\right.\crn
	&&\left.  \hspace{1.8 cm}+m^2_F\fr{a_1a_2}{v_1v_2}\left(-2m^2_1B_0^{(2)}+(m^2_1 +m^2_2)B_0^{(1)}\right)\right]. \label{DfhR} \eea}
 \bibliographystyle{h-physrev}
\bibliography{mainHQTre}
\end{document}